\documentclass[ reprint, amsmath,amssymb,prb]{revtex4-1} 
\usepackage{amsmath}
\usepackage{amsfonts,bbm}
\usepackage{amssymb}
\usepackage{graphicx,color}
\usepackage{subfigure}
\usepackage[normalem]{ulem}

\newcommand{\avg}[1]{\left< #1 \right>}
\newcommand{\bra}[1]{\langle#1|}
\newcommand{\ket}[1]{|#1\rangle}
\newcommand{\braket}[2]{\langle#1|#2\rangle}

\newcommand{\grafe}[1]{\left\{ #1 \right\}}
\newcommand{\tonde}[1]{\left( #1 \right)}
\newcommand{\quadre}[1]{\left[ #1 \right]}

\begin{document}
 \author{Francesca Pietracaprina}
\affiliation{SISSA- International School for Advanced Studies, via Bonomea 265, 34136 Trieste, Italy}
\affiliation{INFN Sezione di Trieste, Via Valerio 2, 34127 Trieste, Italy}
 \author{Valentina Ros}
\affiliation{SISSA- International School for Advanced Studies, via Bonomea 265, 34136 Trieste, Italy}
\affiliation{INFN Sezione di Trieste, Via Valerio 2, 34127 Trieste, Italy}
 \author{Antonello Scardicchio}
 \affiliation{The Abdus Salam ICTP, Strada Costiera 11, 34151 Trieste, Italy}
\affiliation{INFN Sezione di Trieste, Via Valerio 2, 34127 Trieste, Italy}
 \affiliation{Dipartimento di Fisica, Universit\`a degli Studi di Bari ``Aldo Moro'', I-70126, Bari, Italy}
 
\title{The forward approximation as a mean field approximation for the Anderson and Many Body Localization transitions}

\begin{abstract}
In this paper we analyze the predictions of the forward approximation in some models which exhibit an Anderson (single-) or many-body localized phase. This approximation, which consists in summing over the amplitudes of only the shortest paths in the locator expansion, is known to over-estimate the critical value of the disorder which determines the onset of the localized phase. Nevertheless, the results provided by the approximation become more and more accurate as the local coordination (dimensionality) of the graph, defined by the hopping matrix, is made larger. In this sense, the forward approximation can be regarded as a mean field theory for the Anderson transition in infinite dimensions. The sum can be efficiently computed using transfer matrix techniques, and the results are
compared with the most precise exact diagonalization results
available.

For the Anderson problem, we find a critical value of the disorder which is
$0.9\%$ off the most precise available numerical value already in 5
spatial dimensions, while for the many-body localized phase of the
Heisenberg model with random fields the critical
disorder $h_c=4.0\pm 0.3$ is strikingly close to the most recent
results obtained by exact diagonalization. 
In both cases we obtain a critical exponent $\nu=1$. In the Anderson case, the latter does not show dependence on the dimensionality, as it is common within mean field approximations.

We discuss the relevance of the correlations between the shortest paths for both the single- and many-body problems, and comment on the connections of our results with the problem of directed polymers in random medium. 
\end{abstract}

   \maketitle

\section{Introduction}
The propagation of waves and quantum particles in a disordered medium is a fascinating and challenging problem in statistical mechanics, with plenty of relevance for experiments \cite{Abrahams2010,AspectNature2008, InguscioNature2008}. Among the phenomena that occur in such a setting, Anderson localization is probably the most striking \cite{anderson1958absence,Abrahams2010}.  ``Anderson's theorem" in Ref.~\onlinecite{anderson1958absence} states that for sufficiently strong disorder, diffusive transport is completely suppressed in single particle problems on a lattice. The study of the transition that separates the two phases (localized and delocalized) has resisted an exact solution for about 60 years, and numerical methods are still at the core of the advances in this topic \cite{Mir-rev}.

Recently, interest in Anderson localization has surged due to the work by Basko, Aleiner and Altshuler \cite{basko2006metal}, henceforth denoted by BAA. There, the phenomenon of many-body localization (MBL), i.e.\ the stability of the Anderson insulator to the addition of interactions, is investigated. BAA's work has been extended and re-interpreted in several other works \cite{oganesyan2007localization,pal2010mb,de2013ergodicity,serbyn2013local,imbrie2014many,kjall2014many,huse2014phenomenology,laumann2014many,chandran2015constructing,ros2015integrals,luitz2015many}, and MBL appears now to be the most robust mechanism to break the ergodicity that is typical of generic interacting systems. 

The core of BAA's analysis relies on perturbatively accounting for the interactions, at finite temperature and particle density. Technically, they consider the perturbation theory for the imaginary part of the propagator of an excitation on top of an eigenstate by means of the Keldysh formalism. For sufficiently weak interactions, the perturbative series is shown to converge with probability equal to one. In the spirit of Ref.~\onlinecite{anderson1958absence} this implies the localization of the excitations themselves, and the absence of transport.

As an alternative route, the MBL problem can be interpreted as a single particle tight-binding problem in the space of many-body configurations\cite{altshuler1997quasiparticle}, with the interactions playing the role of an effective hopping. However, several issues arise in this formulation. First, the on-site energies in the resulting effective lattice are not independent variables drawn from the same distribution, but they are strongly correlated. Secondly, the connectivity of a configuration in the many-body problem scales with (a power of) the system size: it diverges in the 
thermodynamic limit, and thus it is impossible to define a limiting graph. As a consequence, one needs to define an effective connectivity which stays of $O(1)$ in the thermodynamic limit (a similar phenomenon is observed in Ref.~\onlinecite{huse2014localized}). Finally, the number of paths of a given length connecting two many-body configurations grows factorially with the distance between them, the distance being defined as the minimum number of actions 
of the interaction operator needed to connect the initial to the final configuration. Since the distance between two configurations can be of the order of the system size, the number of paths can grow factorially in the system size. When the mapping to a single particle problem is applied to a system of $N$ interacting spins, it results into a correlated-disorder problem on a (section of) an $N$-dimensional hypercube.

Despite these complications, in some recent works \cite{laumann2014many,ros2015integrals,huse2014localized} this approach has been used successfully to estimate, among other things, the boundaries of the MBL region. The analytical calculations in Refs.~\onlinecite{laumann2014many,huse2014localized} were shown to be in very good agreement with the numerical results in the same papers, obtained with exact diagonalization. The analytic results are derived revisiting an approximation already used in Refs.~\onlinecite{anderson1958absence,abou1973selfconsistent,Medina1992, altshuler1997quasiparticle}, which consists in calculating the Green's functions by retaining only the lowest order in the hopping. Recently, this approximation has been used in Ref.~\onlinecite{de2014anderson} to derive the power-law tail of the distribution of the wave function amplitudes on a Bethe lattice in the localized phase.

In this paper we discuss in detail this approximation, dubbed ``forward approximation" (FA), by illustrating its virtues and limitations and its connections to other, seemingly unrelated, problems of statistical physics. The paper is organized as follows: in Section \ref{sec:derivation} we derive the expression for the wave function amplitudes in forward approximation, and discuss a criterion for localization given in terms of the probability of resonances. In Section \ref{sec:results}, after recalling some known results on the Anderson model on the Bethe lattice, we focus on $d$-dimensional systems (with $d=3-6$). We compare the analytic results we get for the critical disorder within the FA with the numerics, showing how the approximation gives better results as the dimensionality $d$ is increased. We then discuss the application of the aforementioned technique to a many-body problem of interacting spins in a disordered environment. In Section \ref{sec:dompath} we discuss the relevance of the correlations and interference between different paths connecting two configurations, both for the single particle and for the many body problem. In the Conclusions we comment on the various possible directions in which this work can be extended, focusing in particular on the application of the forward approximation to the study of the MBL phase and of the many-body localization-delocalization transition.

\section{Derivation}
\label{sec:derivation}

\subsection{The forward approximation for the eigenfunctions}
\label{par:FA}
To begin with, we derive the expression for the wave function amplitudes in FA for a single particle hopping on a finite lattice with on-site disorder. We consider the Hamiltonian:
\begin{equation}
\label{eq:Hamiltonian}
H=\sum_{i=1}^N\epsilon_i c^\dag_i c_i+t\sum_{\langle i,j \rangle } \left( c^\dag_i c_j+\text{h.c.} \right).
\end{equation}
In the Anderson model, the $\epsilon_i$ are independent random variables uniformly distributed in $[-W/2,W/2]$, one for each of the $N$ sites in the lattice. The edges $\langle i,j \rangle$ define the lattice geometry. The lattice constant is set to one, and we denote with $L$ the length scale characterizing the size of the lattice (for a cubic lattice in dimension $d$ the diameter is $L=N^{1/d}$, for a Bethe lattice or regular random graph it is $L=\ln \left[(N-1)(K-1)/(K+1)+1 \right] /\ln K \approx \ln N/ \ln K$, where $K+1$ is the connectivity of the lattice). The distance $d(a,b)$ between two arbitrary sites $a,b$ of the lattice is the number of edges in a shortest path connecting them. We refer to it as the lattice distance in the following.

We consider the matrix elements of the resolvent between two states associated to the sites $a,b$ in the lattice:
\begin{equation}
G(b,a,E)=\bra{b}\frac{1}{E-H}\ket{a},
\end{equation}
at energy $E$. They have the following expansion in series of $t$
\begin{equation}
\label{eq:path}
G(b,a,E)=\frac{1}{E-\epsilon_a}\sum_{p\in\text{paths}(a,b)}\,\prod_{i\in p}\frac{t}{E-\epsilon_i},
\end{equation}
where the sum runs over all the paths $p$ in the lattice connecting the sites $a$ and $b$, and $i$ labels the sites visited by the path $p$ (site $a$ excluded). Given two sites $a,b$ at lattice distance $n$, the lowest orders in the expansion are 
\begin{equation}
\label{eq:lo}
\begin{split}
G(b,a,E)=&\frac{1}{E-\epsilon_a}\frac{t}{E-\epsilon_1}\frac{t}{E-\epsilon_2}...\frac{t}{E-\epsilon_{n-1}}\frac{t}{E-\epsilon_b}+\\
&+\text{other paths of length $n$}+ \cdots
\end{split}
\end{equation}
where the sites $(a,1,2,...,n-1,b)$ belong to one of the shortest paths connecting $a$ and $b$.

On the other hand, from the spectral decomposition it follows
\begin{equation}
G(b,a,E)=\sum_{\alpha}\frac{\psi_\alpha(b)\psi_\alpha^*(a)}{E-E_\alpha},
\end{equation}
and thus the residue at $E=E_\alpha$ gives 
\begin{equation}
\label{eq:residue}
\lim_{E\to E_\alpha}(E-E_\alpha)G(b,a,E)=\psi_\alpha(b)\psi^*_\alpha(a),
\end{equation}
assuming no degeneracy of the eigenvalues.

The path representation (\ref{eq:path}) does not have a pole at $E_\alpha$, to no order in $t$.  To get the exact poles it is necessary to re-sum the closed paths in the series expansion. Once this is done, the full series is re-cast into a sum over the \emph{non-repeating paths} $\text{paths}^*(a,b)$ connecting the sites $a$ and $b$:
\begin{equation}
\begin{split}
\label{eq:sapath2}
G(b,a,E)=&\frac{1}{E-\epsilon_a- \Sigma_a(E)} \, \times\\
&\sum_{p\in\text{paths}^*(a,b)} \, \prod_{i \in p} \, \frac{t}{E-\epsilon_i-\Sigma^{(p)}_i(E)}.
\end{split}
\end{equation}
Here $\Sigma_a(E)$ is the local self-energy at the site $a$, defined through the identity:
\begin{equation}
\label{eq:defself}
G(a,a,E) \equiv \frac{1}{E- \epsilon_a -\Sigma_a(E)}.
\end{equation}
It is equal to the sum of the amplitudes of all the closed paths in which site $a$ appears only as starting and ending point; to lowest order in $t$
\begin{equation}
\Sigma_a(E)=\sum_{j\in\partial a}\frac{t^2}{E-\epsilon_j}+O(t^3),
\end{equation} 
where $\partial a$ is the set of nearest neighboring sites of $a$. The path-dependent term $\Sigma^{(p)}_i(E)$ is a modified self-energy, which re-sums the loops around site $i$, never crossing site $i$ again, nor any of the sites $(a,1, \cdots, i-1)$ already visited by the non-repeating path $p$. 

The expansion \eqref{eq:sapath2} in non-repeating paths has several advantages. First, while $paths(a,b)$ is an infinite set, even for a finite lattice, $paths^*(a,b)$ is finite for a finite lattice.  Moreover, \eqref{eq:sapath2} is free of the divergences affecting \eqref{eq:path} that are due to \emph{local resonances}, i.e. to sites $i,j$ at bounded distance satisfying $|\epsilon_i-\epsilon_j|\sim \Delta,$ with $\Delta$ small (we clarify the exact meaning of \emph{small} in Sec.~\ref{par:resonances}). Local resonances necessarily occur also in the localized phase, and produce arbitrarily large factors in \eqref{eq:path}, corresponding to the paths repeatedly hitting the resonant sites an arbitrary number of times. These repetitions of large factors lead to the divergence of the naive perturbation series in $t$, but are re-summed into self energy corrections in \eqref{eq:sapath2}. The ``renormalized'' expansion in non-repeating paths is found to converge in the localized phase \cite{anderson1958absence}, when resonances do not proliferate at asymptotically large distances in space, and the hopping hybridizes only the degrees of freedom in a finite, albeit possibly big, region of space. An analogous resummation procedure is discussed in Ref.~\onlinecite{ros2015integrals}, where the perturbation theory for quasi-local conserved operators is shown to converge in the MBL phase.

The expression for the eigenfunction is obtained from the resolvent as follows: the eigenenergy $E_\alpha$ satisfies $E_\alpha=\epsilon_a+\Sigma_a(E_\alpha)$, thus the first factor of \eqref{eq:sapath2} has a pole at $E_\alpha$ with residue $|\psi_\alpha(a)|^2$, as it follows from \eqref{eq:defself} and \eqref{eq:residue}.  Then:
\begin{equation}
\begin{split}
&\lim_{E\to E_\alpha}(E-E_\alpha)G(b,a,E)=\\
&|\psi_\alpha(a)|^2\lim_{E\to E_\alpha}\sum_{p\in\text{paths}^*(a,b)} \, \prod_{i \in p} \, \frac{t}{E-\epsilon_i-\Sigma^{(p)}_i(E)},
\end{split}
\end{equation}
which gives
\begin{equation}
\label{eq:non-repeatingWF}
\psi_\alpha(b)=\psi_\alpha(a)\sum_{p\in\text{paths}^*(a,b)} \, \prod_{i \in p} \, \frac{t}{E_\alpha-\epsilon_i-\Sigma^{(p)}_i(E_\alpha)},
\end{equation}
with $\psi_\alpha(a)$ obtained from $\Sigma_a(E_\alpha)$ using \eqref{eq:residue}.\footnote{A similar argument gives a perturbative series in $t$ for the conductance.}

From \eqref{eq:non-repeatingWF} we can read the expression of the wave function amplitudes to lowest order in $t$. Assume that $\alpha$ labels an eigenstate localized at site $a$ for $t \to 0$. Since $\Sigma_\alpha=O(t^2)$, we have to lowest order $E_\alpha\to\epsilon_a$, $\psi_\alpha(a)\to 1$, $\Sigma_i\to 0$, giving
\begin{equation}
\label{eq:forwardWF}
\psi_\alpha(b)=\sum_{p\in\text{spaths}(a,b)} \, \prod_{i \in p} \, \frac{t}{\epsilon_a-\epsilon_i},
\end{equation}
where the set $\text{spaths}(a,b) \subset \text{paths}^*(a,b)$ contains the shortest paths from $a$ to $b$. Note that this derivation does not rely on the particular structure of the lattice nor on the independence of the on-site energies, and thus it can be straightforwardly generalized to hopping problems on graphs with different geometry or more general distribution of the random energies, such as the ones in Sec.~\ref{par:xxz}.

 The effect of the modified self energy corrections  $\Sigma^{(p)}_i(\epsilon_a)$ is to weaken the role of resonances. Indeed, let us assume that in the path $p$ there is a site $i$ which is resonant with $a$, $|\epsilon_a-\epsilon_i|\sim \Delta$. In this case the forward approximation \eqref{eq:forwardWF} will contain the very large term $t/\Delta$. However, the correction $\Sigma^{(p)}_{i-1}(\epsilon_a)$ in the previous site also contains such large term, leading to a compensation. As we shall see in the following, neglecting this effect leads to an overestimate of the minimum disorder strength needed to localize the system. On the other hand, the loops contributing to the self energy corrections become less relevant when the dimensionality (or connectivity) of the lattice is increased. Thus, the FA is expected to give faithful results in higher dimension (see Sec.~\ref{par:hypercube}).

\subsection{Probability of resonances and criterion for localization}
\label{par:resonances}

For a single particle problem on a finite dimensional lattice, we define an eigenstate $\psi_\alpha$ of a system of size $L$ localized if (with probability 1 over the disorder realizations) the probability of finding a particle at a distance $O(L)$ from the localization center of the state tends to zero in the limit of large $L$. More precisely,
let $a$ denote the localization center of $\psi_\alpha$, and 
\begin{equation}
 \psi_r \equiv \max_{b: \, d(b, a)=r} |\psi_\alpha(b)|.
\end{equation}
We define the state $\psi_\alpha$ localized if there exists a finite $\xi>0$ such that 
\begin{equation}
\label{eq:Ploc}
 P\left(  \frac{\log |\psi_r|^2}{2r}   \leq -  \frac{1}{2\xi}  \right)\to 1 \text{ for } r\to\infty.
\end{equation}
Namely, we require that the random numbers $|\psi_\alpha(b)|^2$ can be enclosed in an exponential envelope for all $b$ sufficiently far from the localization center of the state. By means of the Kubo formula, it is possible to show that this condition on the eigenstates implies the vanishing of the DC conductivity. We identify the localization length of $\psi_\alpha$ with the minimum value of $\xi$ for which Eq.~\eqref{eq:Ploc} is true. It does, in general, depend on the state; however, it is supposed to depend smoothly on the energy $E_\alpha$ in the thermodynamic limit. A mobility edge exists whenever there is band of energies for which such minimum is not finite. In particular, at fixed energy and at the corresponding critical value of disorder $W=W_c$, the localization length diverges and the asymptotic bound in Eq.~\eqref{eq:Ploc} ceases to hold for any finite $\xi$. This entails that for any arbitrarily small, positive $ \epsilon= \xi^{-1}$ and at arbitrarily large distances $r$ from the localization center, there exist sites $b$ such that the ratio $\log |\psi_\alpha(b)|^2/2r$ exceeds the constant $-\epsilon$ with some finite probability.  We expect the delocalized phase $W<W_c$ to be characterized by the stronger condition:
\begin{equation}
 \label{eq:pres}
P\left(  \frac{\log |\psi_r|^2}{2r}   \geq -\epsilon \right)\to 1 \text{ for } r\to\infty,
\end{equation}
holding for any arbitrarily small, strictly positive value of $\epsilon$.

From the equations \eqref{eq:Ploc}, \eqref{eq:pres} it follows that the localization-delocalization transition can be detected analyzing the statistics of the wave function amplitudes as a function of distance. In the following, we compute the wave function amplitudes in FA and determine numerically the  probability in Eq.\eqref{eq:pres}, choosing $\epsilon$ of the order of the numerical precision. We refer to the resulting probability as the ``probability of resonances''. The terminology is motivated by the fact that within the FA, an amplitude of $O(1)$ at a site $b$ at distance $r$ from the localization center $a$ corresponds to a resonance between the two sites. Indeed, the corresponding two sites problem can be considered as a two-level system with reduced Hamiltonian  
\begin{equation}
h=\left( \begin{matrix}
0 & h_r  \\
h_r & \Delta \end{matrix} \right),
\end{equation}
where $\Delta=\epsilon_{a}- \epsilon_b$, and
\begin{equation}
\begin{split}
h_r=
t\mkern-8mu \sum_{p\in\text{paths}^*(a,b)} \prod_{i \in p} \frac{t}{\epsilon_{a}-\epsilon_i-\Sigma_i^{(p)}(\epsilon_{a})},
\end{split}
\end{equation}
where the products are taken over all sites in the path, excluding $a,b$. The sites are resonant when the energy difference $\epsilon_a- \epsilon_b$ is small, and precisely $|\Delta|<h_r $. Considering $h_r$ to lowest order in $t$, one finds that this is equivalent to $|\psi(b)|>1$, with $\psi(b)$ computed in the lowest order FA. Thus, with \eqref{eq:Ploc} and \eqref{eq:pres} one is probing the statistics of resonances within the FA, and requiring that the probability to find at least a resonant site at \emph{any} sufficiently large distance $r$ from the localization center decays to zero in the localized phase. This criterion involving resonances allows to obtain an estimate for the critical disorder within the FA also on the Bethe lattice, where the exact eigenstates satisfy Eq.~\eqref{eq:Ploc} even in the delocalized phase, due to normalization.

\section{Results}
\label{sec:results}
\subsection{Warming up: the Bethe lattice} 
\label{par:Bethe}
The simplest setting for the application of the forward approximation is a Bethe lattice: in this case, given two sites $a$, $b$ there is only one non-repeating path connecting them, along which all the energies are i.i.d.\ . This makes the problem amenable to analytic calculations.\cite{abou1973selfconsistent,Abou-Chacra1974,altshuler1997quasiparticle} We briefly recall some results in the following. 

Let $a$ be the root of the tree, and $K$ the branching number (the connectivity is $K+1$). Within the FA we get that the wave function at one particular point at distance $L$ from the root is given by
\begin{equation}
\psi_L=\prod_{i=1}^L\frac{t}{\epsilon_a-\epsilon_i}.
\end{equation}
 
This random variable has a power-law tail distribution for any distribution of $\epsilon_i$ having a support $S$ such that $\epsilon_a\in S$, as one can see from the divergence of the first moment of the absolute value of $\psi_L$. It is convenient to consider the distribution of the logarithm of $\psi_L$, whose moments are all finite. Let us remind the reader that we choose $\epsilon_i\in[-W/2,W/2]$, and in this calculation for simplicity we set $\epsilon_a=0$. Defining
\begin{equation}
x_L=\ln|\psi_L|^2=-2\sum_i\ln(|\epsilon_i|/t)
\end{equation}
we find that
\begin{equation}
\avg{x_L}=2L\ln(2et/W).
\end{equation}
The ratio $\avg{x_L}/L$ is the typical decay of wave function amplitudes from the origin of localization
\begin{equation}\label{eq:xityp}
\xi_{typ}^{-1}=2\ln(2et/W).
\end{equation}
However, this is \emph{not} the localization length $\xi$ as defined in \eqref{eq:Ploc}. The latter is indeed a uniform bound over the full set of $\sim K^L$ points at distance $L$ from the localization center: it is determined by the decay rate of the maximal amplitude over sites in each shell at distance $L$. This is a point-to-set correlation function decay, which is familiar in the study of disordered systems on the Bethe lattice (on a regular lattice the point-to-set is substituted by the point-to-point, but with exponentially many shortest paths leading to the final point).

The typical value of the maximal amplitude $x^*_L$ among $K^L$ samplings is the largest solution of
\begin{equation}
x^*_L:\quad K^LP(x^*_L)\simeq 1.
\end{equation}
Here $P(x)$ is the distribution of $x_L$, and the $K^L$ paths are treated as independent. For large $L$, what matters is the tail of the distribution $P(x)$. If we rescale
\begin{equation}
z=\frac{x_L}{2L}+\ln(W/2t),
\end{equation}
we get, for large $z$:
\begin{equation}
P(z)\simeq \exp\left(-L(z-1-\ln z)\right).
\end{equation}
The probability distribution of $z$ can be found inverting its Laplace transform, which is computable since $z$ is a sum of i.i.d.\ variables. The maximum $z^*$ over $K^L$ samplings of $z$ is the solution of
\begin{equation}
0=-z^*+1+\ln z^*+\ln K=\ln(z^*eKe^{-z^*}),
\end{equation}
and $x^*_L\simeq -2L\ln(W/2t) + 2 L z^*$. Localization occurs as long as $x^*_L/L<0$, so the critical condition can be written as
\begin{eqnarray}
\ln(Kez^*e^{-z^*})&=&0,\\
z^*-\ln(W_c/2t)&=&0.
\end{eqnarray}
By eliminating $z^*$ we recover the familiar \cite{abou1973selfconsistent} equation:
\begin{equation}
\label{eq:condAbu}
W_c=2teK\ln(W_c/2t).
\end{equation}
Moreover, $z^*$ is given in terms of $W_c$ as
\begin{equation}
z^*=\ln(W_c/2t),
\end{equation}
so we can write the localization length as
\begin{equation}
\xi^{-1}=-x^*_L/L=2\ln(W/W_c).
\end{equation}
This gives a mean-field exponent for the divergence of the localization length at the transition:
\begin{equation}
\xi\simeq\frac{2 W_c}{|W-W_c|}.
\end{equation}

Note that $\xi_{typ}<\xi$ irrespective of the value of the disorder. The belief expressed in Ref. \onlinecite{de2014anderson} is that this behavior persists even within the delocalized regime, in the form of very irregular (multi fractal) eigenstates. 
Note additionally that the difference between $\xi_{typ}$ and $\xi$ is due to the \emph{exponential} sampling of the probability distribution, and it extends both to the finite-dimensional cube and to the many-body case.

\subsection{d-dimensional cube}
\label{par:hypercube}

\begin{figure}
\begin{center}
 \includegraphics[width=.8\linewidth]{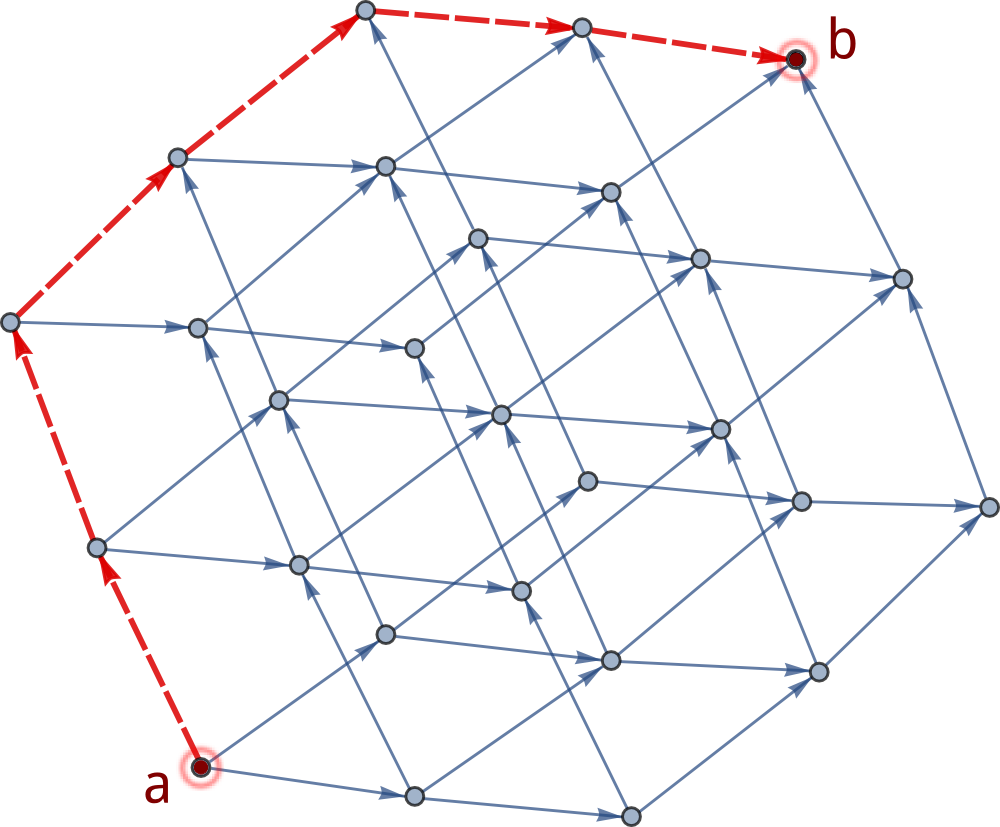}
\caption{(Color online) Anderson model on a cube of side $L=3$. The red, dashed edges form one of the non-repeating paths connecting the sites $a$ and $b$, of length $r=(L-1)d=6$. The other elements in the set $\text{paths}^*(a,b)$ are obtained following the arrows. }\label{fig:hypercube}
\end{center}
\end{figure}

Consider now the case of a $d$-dimensional lattice of side $L$, and let $a$ be the site at one corner of the cube, which we treat as the origin. Given any other site $b$, the orientation of the non-repeating, shortest paths from $a$ to $b$ induces a natural orientation of the edges of the cube, which is thus directed, see Fig.~\ref{fig:hypercube}. Let $r$ be the lattice distance of the sites with respect to the origin $a$: for a cube of side $L$, the maximum $r$ is $r_{max}=(L-1)d$, corresponding to the site at the opposite corner of the cube with respect to $a$. 

In an infinite cube the number of points at distance $r$ from the origin is $(r+d-1)!/((d-1)! r!)$, and thus it grows at most polynomially in $r$, slower than $\sim r^d$. Naively, one would be led to think that the transition is given by the divergence of $\xi_{typ}$, see Eq. \eqref{eq:xityp}. However, the number of minimum length paths leading from the origin to an arbitrary site at lattice distance $r$ scales exponentially with $r$, as $\sim d^r$. Therefore, unlike in the Bethe lattice case, in finite dimension the wave function amplitude at a given site $b$ is a sum over exponentially many correlated terms
\begin{equation}
\label{eq:FAcube}
\psi_\alpha(b)=\sum_{p\in\text{spaths}(a,b)}\prod_{i=1}^r\frac{t}{\epsilon_a-\epsilon_i}= \tonde{\frac{t}{W}}^r \psi'_\alpha(b), 
\end{equation}
where
\begin{equation}
\label{eq:FAcubeRescaled}
 \psi'_\alpha(b)= \sum_{p\in\text{spaths}(a,b)}\prod_{i=1}^r\frac{1}{\epsilon'_a-\epsilon'_i}
\end{equation}
and the random variables $\epsilon'_i$ are uniformly distributed in $\quadre{-1/2, 1/2}$. In the following, we consider the probability distribution of the random variable
\begin{equation}
\label{eq:rescaledZ}
 Z_r \equiv \frac{\log |\psi'_r|^2}{2 r },
\end{equation}
for different values of $r$. Here $\psi'_r$ denotes the maximum among all the rescaled amplitudes \eqref{eq:FAcubeRescaled} at sites that are at lattice distance $r$ with respect to the origin $a$. The probability of resonances for arbitrary values of $t$ and $W$, see Eq.~\eqref{eq:pres}, is easily recovered from the cumulative distribution function of $Z_r$ as: 
\begin{equation}
 \label{eq:presResc}
P\tonde{\frac{\log |\psi_r|^2}{2r} \geq -\epsilon}= P \tonde{Z_r \geq  \log \tonde{\frac{W}{t}}-\epsilon},
\end{equation}
with $\epsilon$ arbitrarily close to zero. According to Eqs.~\eqref{eq:Ploc} and \eqref{eq:pres}, the density of $Z_r$ becomes asymptotically peaked at $\log (W_c/t)$ for $r \to \infty$, with width going to zero with $r$. Thus, the critical value of disorder can be estimated inspecting the scaling with $r$ of the probability density of $Z_r$.

The distribution of $Z_r$ is hard to determine analytically, due to the correlation between the different shortest paths. To account for such correlations, we compute the amplitude \eqref{eq:FAcubeRescaled} numerically by means of a transfer matrix technique, and use the resulting values to determine the probability \eqref{eq:presResc} with $\epsilon$ smaller than the numerical precision. The convenience of the transfer matrix method relies on the fact that it takes only polynomial time in $r$, as it was realized by Medina and Kardar\cite{Medina1992} in their treatment of the Nguyen, Spivak, and Shklovskii \cite{nguyen1985pis, *nguyen1985jetp2,*nguyen1986hopping} (NSS) model. 
\footnote{Note that the present calculation does not reduce to the one for the NSS case, since a major difference between the two models is in the statistics of energy denominators: in the NSS model the binary disorder $\epsilon_i=\pm W$ (with probability $p$ or $1-p$) does not allow for resonances due to a single site. Rather, the resonances arise from contributions of different paths. This led to a body of work following 
Ref.~\onlinecite{Medina1992}, on the presence of a sign transition, where effects from different paths accumulate in order to break the sign symmetry. In the FA for the Anderson case, the energy denominators can be arbitrarily small with finite probability, generating path weights that are fat tailed distributed. The fat tail of this distribution is crucial for the considerations below.}

The numerical computation is as follows. We fix $t=1$ and introduce the matrix $\mathcal T$ defined as
\begin{equation}
\mathcal T= \mathcal W A_f,
\end{equation}
where $A_f$ is the forward adjacency matrix of the lattice (that is, the adjacency matrix associated to the directed cube of Fig.\ref{fig:hypercube}), and $\mathcal W$ is a diagonal matrix with components:
\begin{equation}
\mathcal W=\text{diag}\left(\frac{1}{\epsilon'_a-\epsilon'_k}\right)_{k=1,..,L^d}.
\end{equation}

We initialize the system in the state $\ket{\psi^{(0)}}=|a\rangle$ completely localized in the origin $a$, and iteratively apply the transfer matrix $\mathcal T$. A single iteration gives 
\begin{equation}
 \ket{\psi^{(1)}} \equiv \mathcal T \ket{\psi^{(0)}}= \frac{1}{\epsilon'_a- \epsilon'_{l_1}}\ket{l_1} + \frac{1}{\epsilon'_a- \epsilon'_{l_2}}\ket{l_2}+ \dots, 
\end{equation}
where $l_1, \cdots, l_d$ are the forward neighbors of site $a$. 
The value of $\psi_\alpha(b)$ equals $\psi_\alpha(b)=\braket{b}{\psi^{(r)}}$, where $\ket{b}$ is the state completely localized in the site $b$ and $r$ is the lattice distance between $a$ and $b$.

We fix $\epsilon'_a=0$ and compute the rescaled amplitude
\eqref{eq:FAcubeRescaled} for all the points $b$ on a shell at {the same} lattice distance $r=r_{max}-c$ from the origin of a hypercube of side $L$.  Here $c\sim O(1)$ is fixed so as to have about $20$ points per each size of the hypercube. We determine the maximal $\psi'_r$ among the wave function amplitudes on those sites. We repeat the procedure for hypercubes of different sizes, with $O(10^5)$ disorder realizations for most system sizes, decreasing to $O(10^3)$ realizations only for the biggest system sizes that we consider (e.g. in $d=3$ we take system sizes $r=10$ through $292$, with $1.5\cdot 10^5$ disorder realizations up to $r=202$ and $2.5\cdot 10^3$ realizations up to $r=292$).

As we discuss in Section \ref{sec:dompath}, the main contribution to  the transfer matrix result comes from only one of the exponentially many paths in \eqref{eq:FAcubeRescaled}, and the results obtained with the transfer matrix technique are faithfully reproduced by analyzing the statistics of the dominant path alone. The latter can be determined (see Section \ref{sec:dompath}) with an algorithm that is computationally more efficient than the transfer matrix, allowing to access to much bigger system sizes. The results presented in this sections for $d=6,7$, as well as for the higher values of $r$ in $d=3-5$, are obtained with this procedure.

\subsubsection{Fluctuations of the wave function amplitudes}

\begin{figure}
\begin{center}
 \includegraphics[width = \columnwidth]{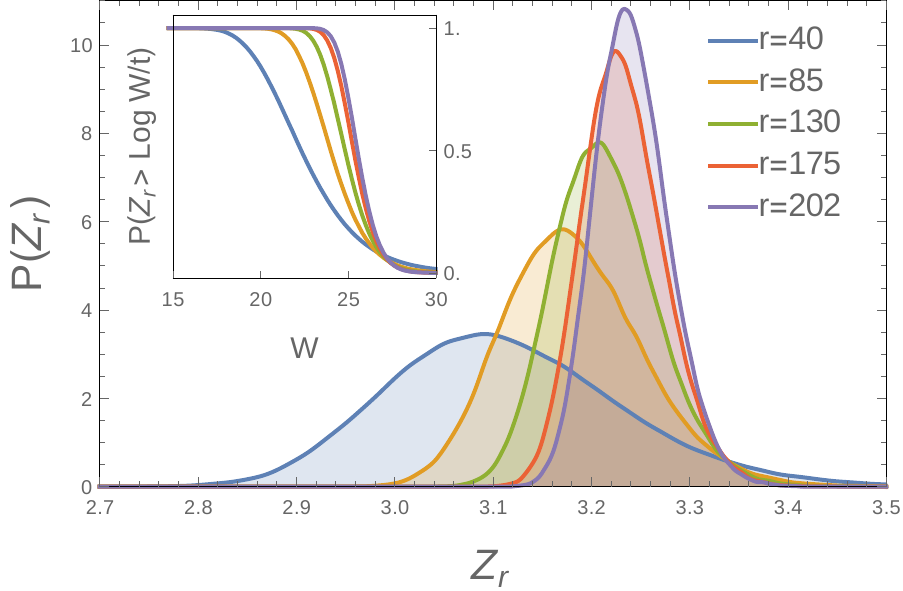}
\caption{(Color online) Probability density of the variable $Z_r$ defined in Eq.~\eqref{eq:rescaledZ}, for different $r$ and $d=3$. For $r\to \infty$, the curves become peaked around the critical value $\log (W_c/t)$. Inset: cumulative distribution function. Each curve is obtained with $1.5 \cdot 10^5$ disorder realizations. Very similar results are obtained for higher dimensionality.}\label{fig:PlaceHolderPZ}
\end{center}
\end{figure}

\begin{figure}[htbp]
\begin{center}
 \includegraphics[width = \columnwidth]{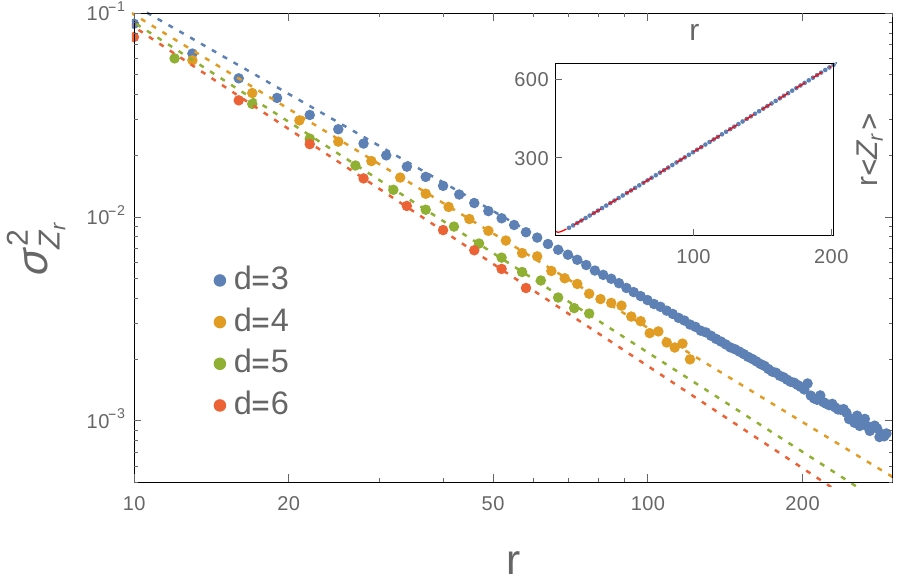}
\caption{(Color online) Variance $\sigma^2_{Z_r}$ of the variable $Z_r$ defined in Eq.~\eqref{eq:rescaledZ}. The plot is in log-log scale. The points corresponding to larger $r$ are fitted linearly, according to the scaling form Eq.~\eqref{eq:ansatz}, and the values of the exponents $\omega_{FA}(d)$ reported in Table~\ref{tab:omega} are extracted from the coefficient of the linear term in the fit. The number of realizations is $1.5 \cdot 10^5$ for $r$ smaller than $202$, $53$, $52$, $40$  for $d=3$, $4$, $5$, and $6$, respectively, and $2 \cdot 10^3$ for larger values of $r$. \textit{Inset} Mean value of the variable $ r Z_r $ defined in Eq.~\eqref{eq:rescaledZ}, for $d=3$. The fit is linear with a correction $\propto r^{\omega_{FA}(3)}$, in agreement with the scaling form in Eq.~\eqref{eq:ansatz}, with the value of $\omega_{FA}(3)$ given in Table \ref{tab:omega}. The results of the fit are, with reference to Eq. \eqref{eq:meanZfit}: $c_1=-18.2\pm0.3$, $W_c=27.03\pm0.02$, $c_2=29.6\pm0.8$. The same behavior holds for higher dimensionality and results in the estimates of the critical disorder values in Table \ref{default}. }\label{fig:variancescaling}
\end{center}
\end{figure}

\begin{figure}[htbp]
\begin{center}
 \includegraphics[width = \columnwidth]{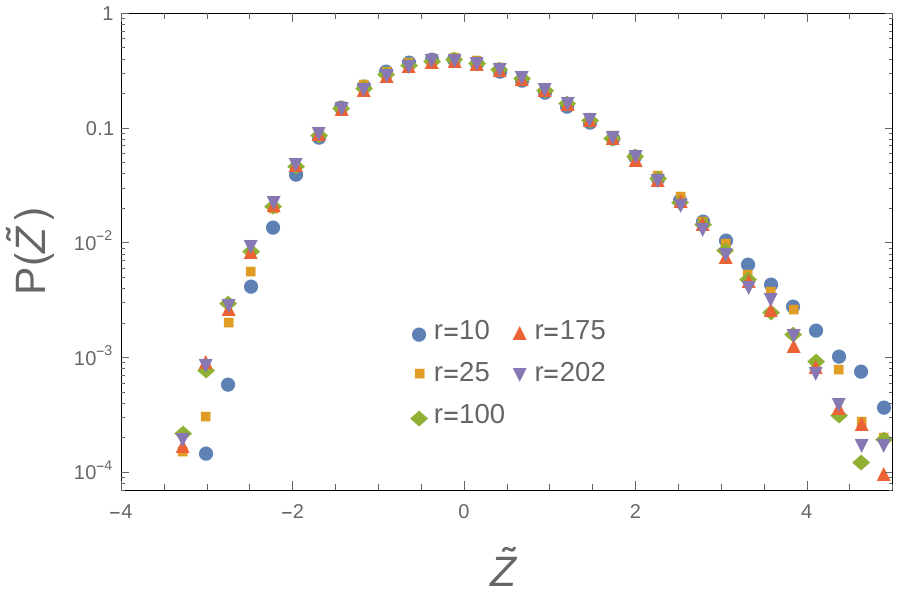}
   \includegraphics[width = \columnwidth]{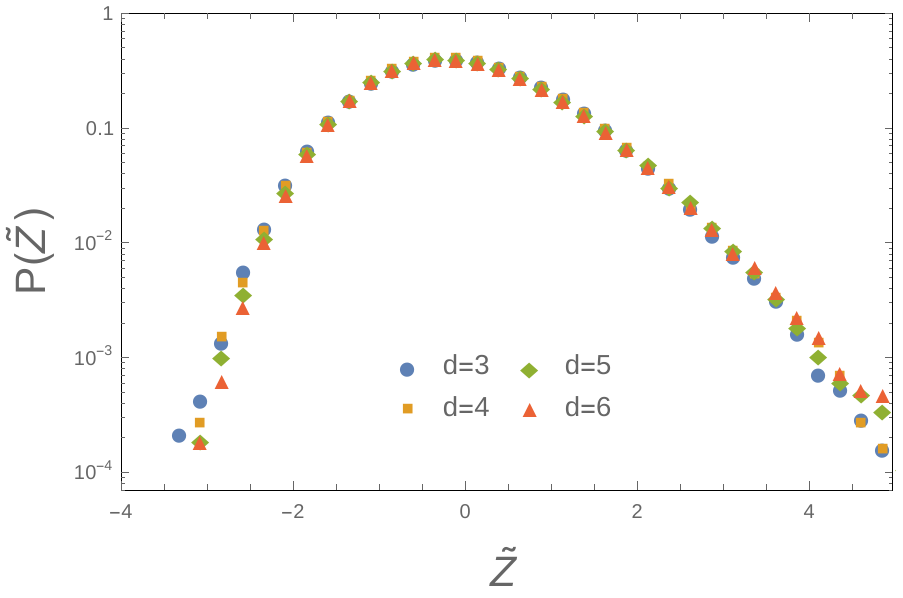}
     
   \caption{(Color online) Probability density $P(\tilde Z)$ of the variable $\tilde{Z}_r$ defined in Eq.~\eqref{eq:Ztilde}. Each curve is obtained with $1.5 \cdot 10^5$ realizations.  {\it Top.} Density of  $\tilde{Z}_r$ for different values $r$ and $d=3$. The curves seem to converge to a unique limiting distribution with increasing $r$. {\it Bottom.} Density of  $\tilde{Z}_r$  for fixed $r=52$ and different dimensionalities. }\label{fig:collapse}
\end{center}
\end{figure}

In Fig.~\ref{fig:PlaceHolderPZ} we plot the probability density of the variable $Z_r$ defined in Eq.~\eqref{eq:rescaledZ}, for different values of $r$ in $d=3$. The plot shows a drift of the position of the peaks with increasing $r$, together with the shrinking of the width of the distribution, in agreement with the conditions \eqref{eq:Ploc}, \eqref{eq:pres}. Plots of the $r$-dependence of the variance $\sigma^2_{Z_r}$ of \eqref{eq:rescaledZ} are given in Fig.~\ref{fig:variancescaling}, in log-log scale for $d=3-6$. The linear behavior indicates that the fluctuations of $Z_r$ decay to zero as a power law in $r$, with a coefficient that depends on the dimensionality. The higher cumulants of the distribution exhibit a similar linear behavior in log-log scale. Moreover, the numerical computation indicates that for fixed $d$ the probability densities of the variable 
 \begin{equation}\label{eq:Ztilde}
  \tilde Z_r=\frac{Z_r-\langle Z_r\rangle}{\sigma_{Z_r}}
 \end{equation}
collapse to a limiting curve for increasing $r$, see Fig.~\ref{fig:collapse}. As shown in the same plot, for fixed $r$ and varying dimensionality, the distribution of $\tilde{Z}_r$ does not change significantly, except for a weak $d$-dependence of the tails.

These numerical observations are compatible with the following large $r$ scaling form for $Z_r$:
\begin{equation}
\label{eq:ansatz}
r Z_r \underset{r \to \infty}{\sim} r \log \tonde{\frac{W_c}{t}} + r^{\omega(d)} u,
\end{equation}
where $u$ is a random variable of $O(1)$ with a distribution which depends weakly on the dimensionality.

\begin{table}[htdp]
\caption{Values of the exponent $\omega_{FA}(d)$ governing the decay of the fluctuations of $Z_r$ with $r$, see Eq.~(\ref{eq:ansatz}). A comparison is made with the values of the droplet exponents $\omega_{DP}(D)$ obtained numerically for the directed polymer in dimension $1+ (d-1)$. The numerical values are taken from Appendix A in Ref.~\onlinecite{monthus2012RandomIsingHighD}.}
\begin{center}
\begin{tabular}{c|c|c}
d=D+1 & $\omega_{FA}(d)$ & $\omega_{DP}(D)$\\
\hline 
\hline
3 & $0.278 \pm0.005$ & $0.244$\\
4 & $0.23\pm0.01$ & $0.186$\\
5 & $0.191 \pm 0.007$ & $0.153$\\
6 &  $0.168 \pm 0.006$ & $0.130$
\end{tabular}
\end{center}
\label{tab:omega}
\end{table}%

According to \eqref{eq:ansatz}, for large $r$ the fluctuations $ \sigma^2_{Z_r}$ decay to zero with the power $r^{2(\omega(d)-1)}$. From the linear fit of $\log \tonde{\sigma^2_{Z_r}} $ we extract the numerical estimate of the exponent in \eqref{eq:ansatz}, which we denote with $\omega_{FA}(d)$. The results are reported in Table~\ref{tab:omega}. 

\begin{figure}
\begin{center}
 \includegraphics[width = \columnwidth]{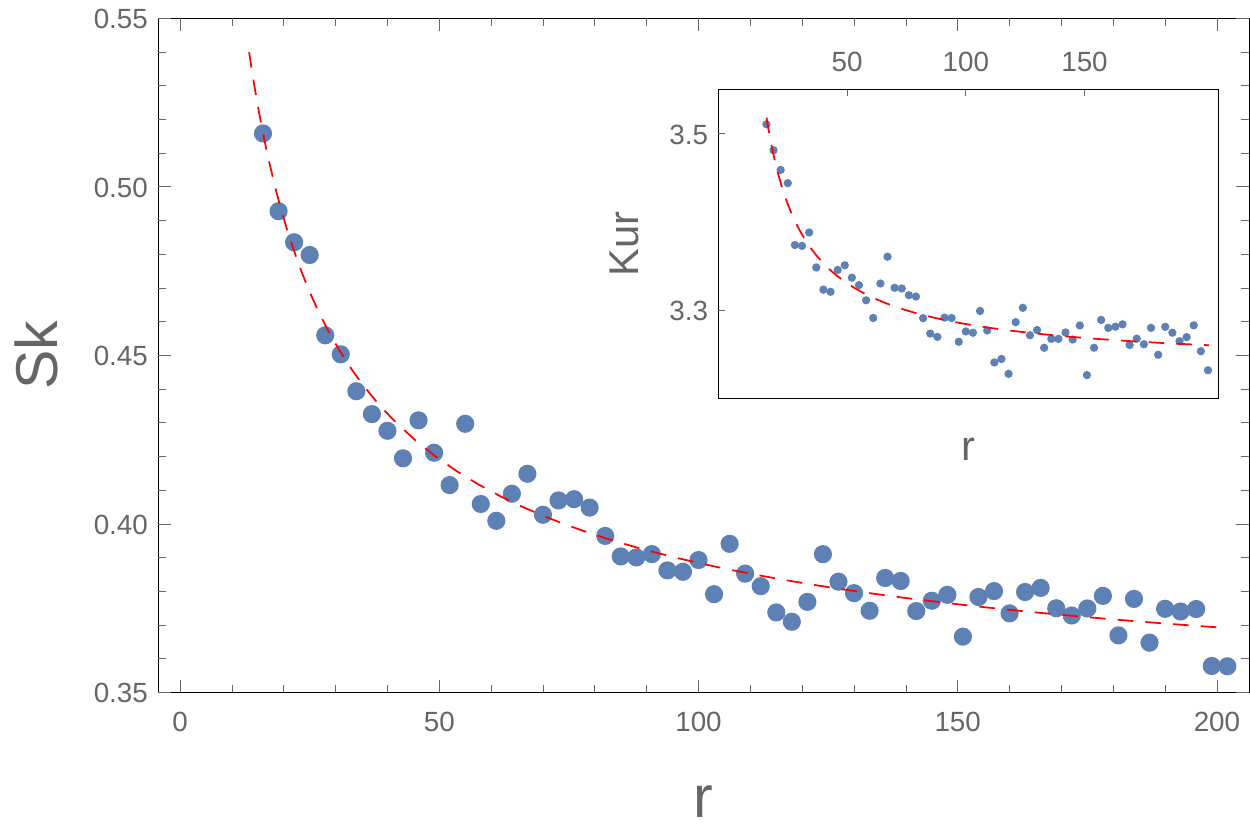}
\caption{(Color online) Skewness $\text{Sk}$ of the distribution of the variable $\tilde{Z_r}$ defined in Eq.~\eqref{eq:Ztilde}, for $d=3$. The red dashed line is a fit of the form $\alpha+ \beta r^{\gamma}$, with  $\alpha, \beta, \gamma$ free parameters. The coefficient $\alpha$ is the estimate of the asymptotic value of the skewness, and it equals $\text{Sk} =0.34 \pm 0.02$.  \textit{Inset. } Kurtosis $\text{Kur}$ of the distribution of $\tilde{Z_r}$ for $d=3$, as a function of $r$. The fitting procedure is analogous to the one for the skewness, and results in $\text{Kur} =3.24 \pm 0.04$.}
\label{fig:skewness}
\end{center}
\end{figure}

In order to characterize the limiting distribution in  Fig.~\ref{fig:collapse}, we compute the skewness $\text{Sk}= \kappa_3/\kappa_2^{3/2}$ and the kurtosis $\text{Kur}= \kappa_4/ \kappa_2^2$  of the density of $\tilde{Z}_r$ (here $\kappa_i$ denotes the $i$-th cumulant of the distribution). From \eqref{eq:ansatz} it follows that these parameters approach the ones corresponding to the variable $u$ in the limit of large $r$.  We restrict to $d=3$, for which we have the largest statistics available. Plots of the $r$-dependence of $\text{Sk}$ and $\text{Kur}$ are given in Fig.~\ref{fig:skewness}. The asymptotic values are estimated to be $\text{Sk}=0.34 \pm 0.02$ and $\text{Kur}=3.24 \pm 0.04$, see the caption of Fig.~\ref{fig:skewness} for details.

\subsubsection{Estimate of the critical disorder}
To determine the critical value of disorder for $t=1$, we  extrapolate the asymptotic limit of the typical value of $Z_r$. Since the distribution is not fat tailed, we can equivalently consider the averages of $Z_r$ and set: 
\begin{equation}
 \langle Z_\infty\rangle \equiv \lim_{r \to \infty} \langle Z_r \rangle=\log\tonde{W_c}.
\end{equation}
The inset in Fig.~\ref{fig:variancescaling} shows the scaling with $r$ of $ r \langle Z_r \rangle$. The average grows linearly in $r$, in agreement with Eq.~\eqref{eq:ansatz}. We fit the data with the form
\begin{equation}
 \langle r Z_r\rangle=c_1+\,\log (W_c)\, r+c_2 \,r^{\omega(d)},
 \label{eq:meanZfit}
\end{equation}
with the numerical values $\omega(d)= \omega_{FA}(d)$ reported in Table ~\ref{tab:omega},.

The resulting estimates of the critical disorder, which we denote with $W_c^{FA}$, are displayed in Table \ref{default}. For the smallest dimensions, a comparison is made with the critical values $W_c^{\text{num}}$ determined in Refs.~\onlinecite{slevin2014critical, slevin2014dimensional} by means of a combination of exact diagonalization and transfer matrix techniques.

\begin{table}[htdp]
\caption{Comparison between the critical value for localization in the Anderson model in $d$ dimensions predicted by the forward approximation ($W^{FA}_c$) and the numerical results ($W^{\text{num}}_c$) of Ref.~\onlinecite{slevin2014critical}. The relative error decreases faster than $d^{-6}$, presumably exponentially. For $d=6$ the transition value $W_c^{FA}=77.0\pm0.3$ can be compared with the result of Ref.~\onlinecite{Cuevas2007}, $W_c^{d=6}=74.5\pm0.7$. This number is however an underestimation of the transition due to the choice of boundary conditions. For 7 dimensions there is no available numerics to compare with.}
\begin{center}
\begin{tabular}{c|c|c|c}
d & $W_c^{FA}$ & $W_c^{\text{num}}$& Error\\
\hline
\hline
3 & $27.03 \pm0.03$ & $16.536 \pm 0.007$&39\%\\
4 & $41.4\pm0.1$ & $34.62 \pm 0.03$&16\%\\
5 & $57.8 \pm 0.2$ & $57.30 \pm 0.05$&0.9\%\\
6 &  $77.0 \pm 0.3$ & -&-\\
7 &  $93.8 \pm 0.3$ & -&-

\end{tabular}
\end{center}
\label{default}
\end{table}%

The data in Table \ref{default} clearly show that the FA gives an upper bound to the critical disorder, since the renormalization of the energy denominators provided by the (modified) self-energy corrections are neglected, and the effects of resonances are thus enhanced. However, increasing the dimensionality the discrepancy between the numerical estimates of $W_c$ decreases; the enhanced precision of the FA result is due to the fact that the loops giving rise to the self-energy corrections become less relevant in higher dimensional lattices, and thus the FA becomes asymptotically exact in this limit.

\subsubsection{Divergent length scales and critical exponents}

For fixed values of $W$ and for finite $r$, the probability of resonances \eqref{eq:presResc} is determined by the tails of the distribution of $Z_r$. 

For increasing $r$, the asymptotic limit is approached in a different way at the two sides of the transition: for $W>W_c$, the probability of resonances goes to zero exponentially with $r$. Below the transition, the probability converges to one much faster, with corrections that are only double exponential in $r$. We justify analytically this behavior in Appendix \ref{appendix:ProbDen}, by computing an approximate expression for the density of the variable $Z_r$. The approximation consists in considering the different paths contributing to it as independent variables.

\begin{figure}
\begin{center}
 \includegraphics[width = \columnwidth]{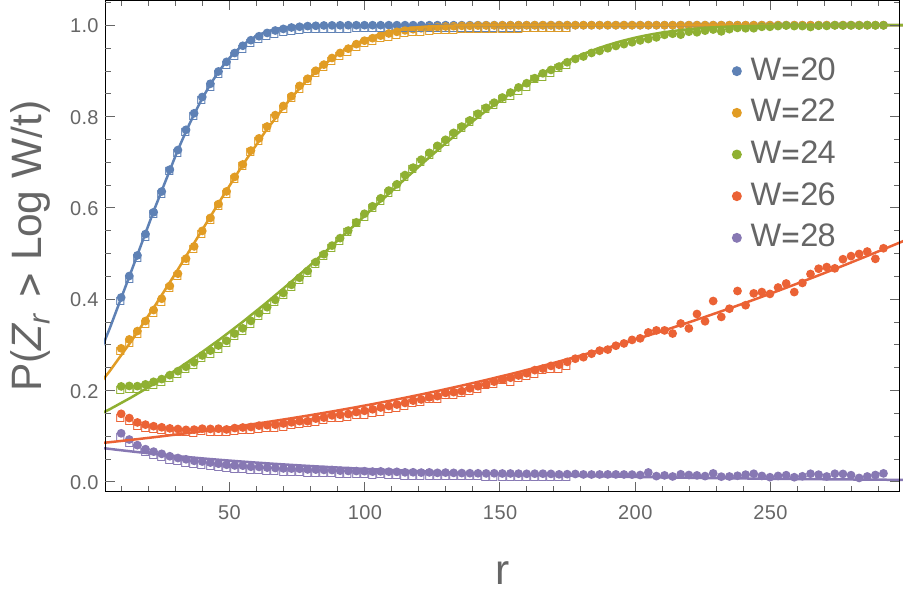}
\caption{(Color online) Probability of resonances $P\tonde{Z_r>\log (W/t)}$ for the variable $Z_r$ defined in Eq.~\eqref{eq:Ztilde} and $d=3$.  Asymptotically in $r$, the probability reaches $0$ exponentially fast in the localized phase, and it reaches $1$ double exponentially fast in the delocalized phase, in agreement with the analytic computations in Appendix \ref{appendix:ProbDen}. In the plot, the squares are the results of the transfer matrix calculation, the points of  the dominating path (see Sec.~\ref{sec:dompath}) while the continuous lines are the exponential or double exponential fits. Very similar results are obtained for higher dimensionality.}\label{fig:PLfit}
\end{center}
\end{figure}

Examples of the fits of the probability of resonances are shown in Fig.~\ref{fig:PLfit}.
To extract a $W$-dependent length scale $l(W)$, we perform an exponential fit of the form:
\begin{equation}\label{eq:fitPloc}
 \begin{split}
  P\tonde{\frac{\log |\psi_r|^2}{2r}>0}= a_1(W) \text{exp}\quadre{-\frac{r}{l(W)}}
 \end{split}
\end{equation}
for $W>W_c$. For $W<W_c$ we determine $l(W)$ by means of the linear fit
\begin{equation}\label{eq:fitPdeloc}
 \begin{split}
\log \left|\log \quadre{1-P\tonde{\frac{\log |\psi_r|^2}{2r}>0}}\right|= a_{2}(W)- \frac{r}{l(W)}.
 \end{split}
\end{equation}

\begin{figure}
\begin{center}
 \includegraphics[width = \columnwidth]{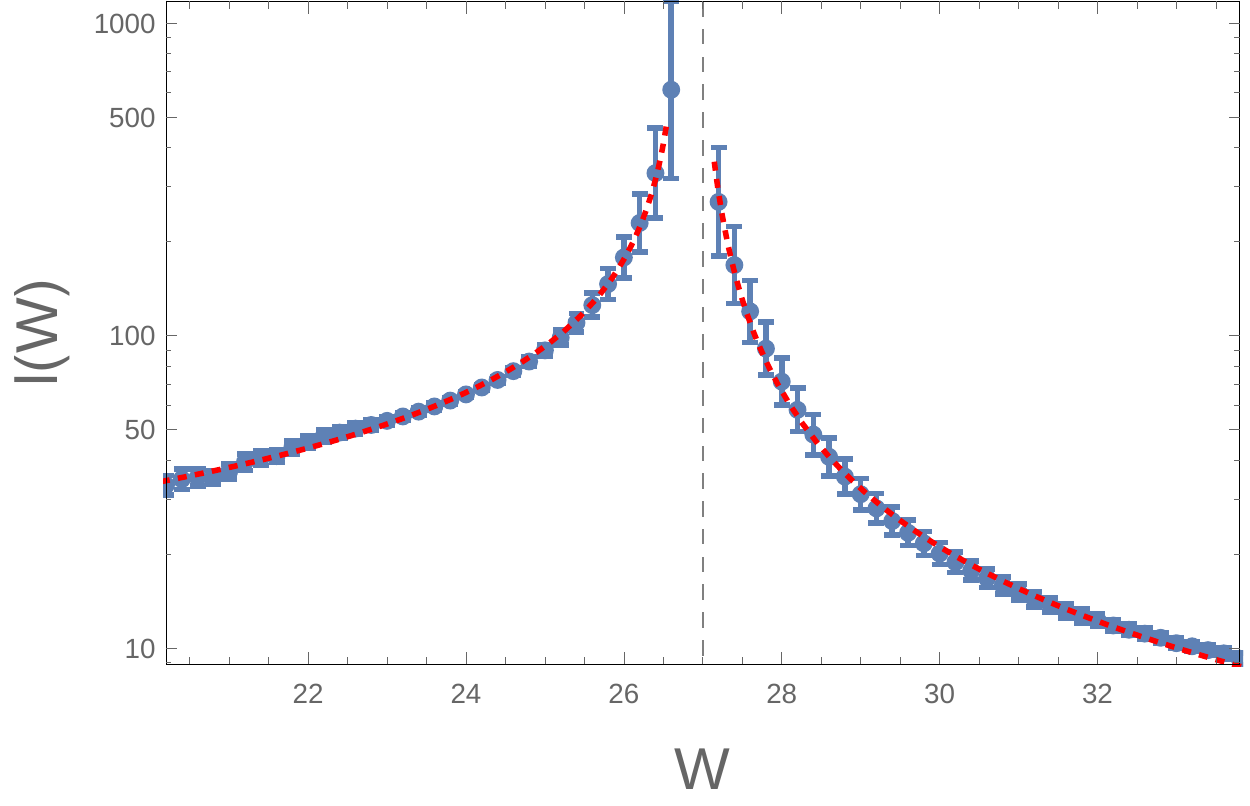}
\caption{ (Color online) Power law divergence of the length scale $l(W)$ defined in Eqs.~\eqref{eq:fitPloc} and \eqref{eq:fitPdeloc}. The values of $l(W)$ for fixed $W$ are determined from fits such as the ones in Fig.~\ref{fig:PLfit}. The power law fit produces a critical exponent $\nu \simeq 1$ and a critical value $W_c$ compatible with the ones listed in Table \ref{default}. The results shown here are for $d=3$; very similar results are obtained for higher dimensionality. Notice how in the delocalized phase the distance to observe a resonance is typically larger (for the same $|W-W_c|$) than the localization length in the localized phase.}
\label{fig:nu3}
\end{center}
\end{figure}

The length scale $l(W)$ is plotted in Fig.~\ref{fig:nu3} for $d=3$. We expect it to diverge in the same way as the localization length/correlation length does in the localized/delocalized phase, respectively. 
 We find that $l(W)$ diverges as a power-law at a critical disorder compatible with the values of $W_c^{FA}$ listed in Table \ref{default}. A fit of the form $\log \tonde{ l(W)}=\log c-\nu\log|W-W_c^{FA}|$ results in an exponent that is compatible with $\nu\approx1$ for all dimensions, consistently with the Bethe lattice picture and with the results in Appendix \ref{appendix:ProbDen}. However, some deviations can be observed: a more careful analysis of the numerical data will be presented in a future publication.

\subsubsection{Connections with the problem of directed polymers in random medium}
\label{sub:polymer}

In the single particle case, the energy denominators associated to different sites along the paths are independent variables. Thus, the expression for the wave function amplitude in FA, Eq. \eqref{eq:forwardWF}, resembles the expression for the partition function of a directed polymer (DP) in a random potential\cite{DerridaPolymerProceeding, deridda1988polymer, halpin1995kinetic}, with the thermal weights for the polymer configurations given by the amplitudes of the different paths. This analogy is not straightforward, due to the occurrence of negative contributions in \eqref{eq:forwardWF}. Nevertheless, it has been fruitfully exploited both for the single particle problem \cite{Somoza2007, Somoza2009, SomozaLeDoussal2015, MonthusGarel2009Statistics}, and for problems of interacting spins on the Bethe lattice \cite{Feigelman2010, IoffeMezard2010, Mueller2013, Yu2013}. 

Motivated by this analogy ,the authors of Ref.~\onlinecite{Somoza2009} have proposed a scaling form analogous to \eqref{eq:ansatz} for the logarithm $\log g$ of the conductance of an Anderson model. There, the conductance in $d=2$ is obtained from the Green functions, which are computed numerically within a modified FA, the modification consisting in taking energy denominators that are not arbitrarily small but are bounded from below. \footnote{Note that if this constraint is relaxed and the energy denominators are allowed to be arbitrarily small, $\log g$ is found to be proportional to the quantity $r Z_r$ that we are considering.} It is shown that the fluctuations of $\log g$ scale with an exponent $\omega(d=2)=1/3$, and that the distribution of the variable $u$ is compatible with a Tracy-Widom distribution. 

These results are consistent with the conjecture\cite{Medina1989, Medina1992} that in the strongly localized phase, where the expansion in non-repeating paths is best controlled, the Anderson model in dimension $d$ belongs to the same universality class of the directed polymer in dimension $1+ D$, with $D=d-1$. In particular, the conjecture implies that in the limit of large $r$ the distribution of $ \log g$ has the scaling form \eqref{eq:ansatz}, with $\omega(d)$ coinciding with the droplet exponent\cite{FisherHuse1991} in $1+(d-1)$ dimensions (which is exactly known\cite{Huse1985} to be equal to $1/3$ for $D=1$), and $u$ having the same distribution of the fluctuations of the free energy in the disordered phase of the polymer (distributed according to the Tracy-Widom distribution\cite{Johansson2000, Prahofer2000, *prahofer2002scale, *prahofer2004exact} in $D=1$).

The values of the scaling exponents extracted from our data do not compare well with the droplet exponents $\omega(D=d-1)$ of the DP, see Table~\ref{tab:omega}.
Curiously, they compare within errors with $\omega(D+1)$. We do not have an explanation for this curious behavior, and we leave its analysis for future work. Broadly speaking, the discrepancies with respect to the directed polymer results are generated by the fat-tail of the distribution of the paths amplitudes in \eqref{eq:FAcube}, produced by the arbitrarily small energy denominators. It might be that the finite size effect are more pronounced in the case of unbounded denominators. On the other hand, it is quite natural to expect that the models of non-repeating paths with bounded amplitude considered in Ref.~\onlinecite{Somoza2009} exhibit a stronger dependence on the dimensionality, due to the fact that the domination by one single path is less pronounced in that case. We comment more on this point in Sec.~\ref{sec:dompath}.

\subsection{Heisenberg model with random fields}
\label{par:xxz}

\begin{figure}
\begin{center}
 \includegraphics[width = \columnwidth]{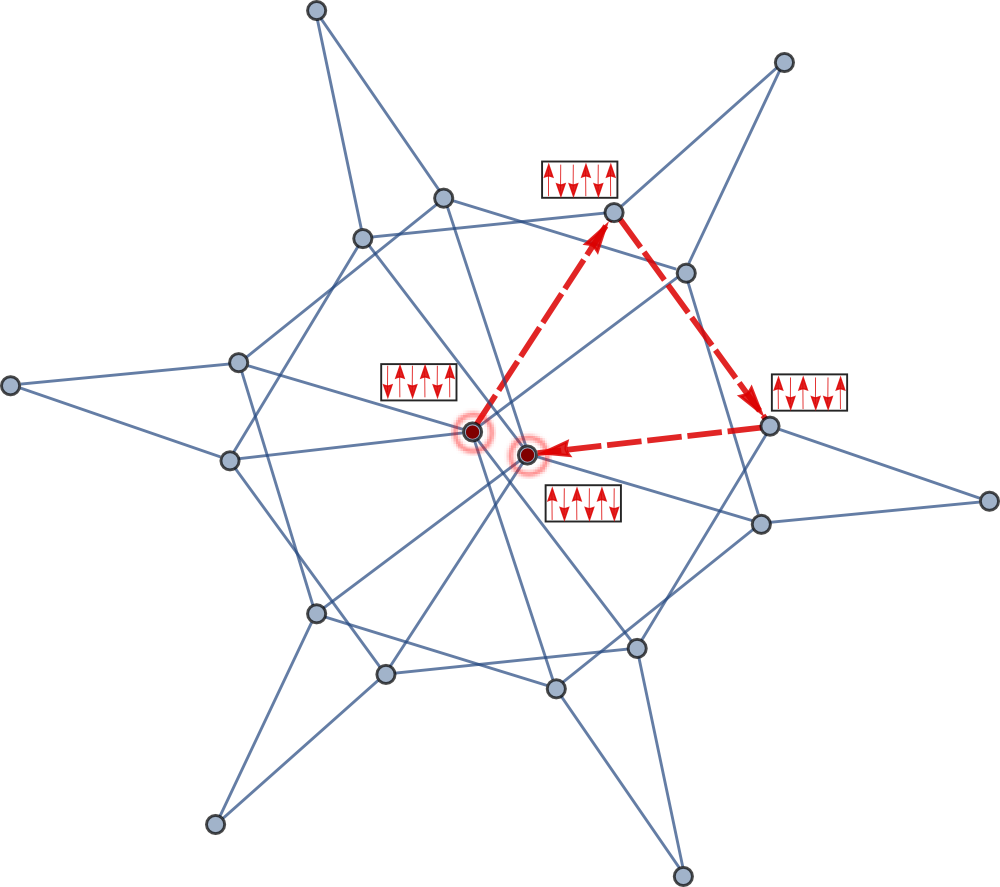}
\caption{(Color online) Graph corresponding to the configuration space of the XXZ spin chain, see Eq.~\eqref{eq:HamXXZ}, of total length $L=6$ and with periodic boundary conditions. Each site in the graph is associated to a product state in the basis of $s^z_i$ operators. Only sites corresponding to states with zero total spin are represented. The initial Neel state  $|\downarrow\uparrow\dots\rangle$ and the final, totally flipped state  $|\uparrow\downarrow\dots\rangle$ are highlighted with circles. The red, dashed edges form one of the shortest paths connecting the two states, of length $L/2=3$. }\label{fig:xxzgraph}
\end{center}
\end{figure}

In order to test the forward approximation on a many-body problem, we consider an XXZ spin-1/2 chain in random magnetic field,
\begin{equation}
\label{eq:HamXXZ}
H(t)=-\sum_{i=1}^L h_i s_i^z-\Delta\sum_{i=1}^Ls^z_is^z_{i+1}-t\sum_{i=1}^L(s^x_is^x_{i+1}+s^y_is^y_{i+1}),
\end{equation}
where periodic boundary conditions are assumed ($s^\alpha_1=s^\alpha_{L+1}$), and the random fields $h_i$ are uniformly distributed in $\quadre{-h, h}$. This spin Hamiltonian \eqref{eq:HamXXZ} has been studied in a large number of works \cite{pal2010mb, znidaric2008many, Bardason2012, de2013ergodicity, Bauer:2013rw, nanduri2014entanglement,Reichman2014absence,mondragon2015, Goold2015totalcorrelations}, in which numerical evidence of the existence of a localization-delocalization transition is provided, mainly based on exact diagonalization results. The critical disorder is estimated \cite{luitz2015many} to be $ h_c \simeq 3.72(6)$ for states in the middle of the energy band  and parameters $t=1$ and $\Delta=1$.

As mentioned in the Introduction, the many body problem can be seen as a single particle hopping problem in the ``configuration space". The latter is composed of the $2^L$ product states in the basis of $s^z_i$, which span the full Hilbert space and diagonalize $H(0)$. We denote these basis states with $|n \rangle $, and refer to them as the ``computational basis''. The mapping to an hopping problem is obtained by interpreting each state $|n \rangle $ as a vertex $n$ of a graph, with associated random energy $E_n$ defined by $H(0) | n \rangle= E_n |n \rangle$. The third term in \eqref{eq:HamXXZ} provides the hopping between different sites, thus defining the geometry of the graph. Note that due to spin conservation, the full configuration space, and consequently the graph, are partitioned into disjoint sectors corresponding to different total spin; we restrict to the sector of total spin equal to zero, corresponding to a connected graph with $\binom{L}{L/2}$ vertices. See Fig.~\ref{fig:xxzgraph} for a pictorial representation of the graph for $L=6$.

The effective hopping problem can be analyzed using the procedure set up in Sec.~\ref{par:FA}: the amplitude $\Psi_\alpha$ of an eigenstate of the effective single particle {problem} is given in forward approximation by:
\begin{equation}
\label{eq:FAxxz}
\Psi_\alpha(n_2)=\sum_{p\in\text{spaths}(n_1,n_2)}\prod_{n \in p}\frac{t}{E_{n_1}-E_n},
\end{equation}
where it is assumed that the eigenstate satisfies $\Psi_\alpha(n) \to \delta_{n, n_1}$ for $t \to 0$.

In the many body language, Eq.(\ref{eq:FAxxz}) provides the expression of the coefficients of the eigenstates of \eqref{eq:HamXXZ} in the computational basis, to lowest order in the coupling $t$. The exponential decay of the coefficients implies localization in the configuration space, meaning that the full many-body eigenstates are effectively a superposition of product states which differ only by configurations distant $O(1)$ from the initial configuration. The two main consequences of this structure of the wave function is that they have significantly less entanglement than ergodic states \cite{buccheri2011structure,de2013ergodicity,Bauer:2013rw} and that, using Kubo's formula for linear response \cite{ros2015integrals}, one can prove that they cannot support transport on macroscopic distances.

Similarly to the Anderson case, we fix an initial configuration of spins and we look at the amplitude in perturbation theory on the most distant, fully flipped configuration. In particular, we fix the localization center to be the site correspondent to the Neel state $|n_1 \rangle=|\downarrow\uparrow\dots\rangle$, and consider the wave function amplitude on the site corresponding to the fully flipped Neel state $|n_2\rangle=|\uparrow\downarrow\dots\rangle$. These two sites $n_1, n_2$ are connected by $2(L/2)!$ paths on the graph, of length $r=L/2$ each.

By means of the transfer matrix we compute the rescaled amplitude
\begin{equation}
\label{eq:Zrh}
 Z_r(h)\equiv\frac{\log|\Psi_r|^2}{2r}
\end{equation}
for different disorder strength $h$, with $\Psi_r$ given by \eqref{eq:FAxxz}. We consider spin chains of size $6-20$ with hopping and interaction constants respectively $t=1$ and $\Delta=1$, and $h=1-6$. Note that, despite the general framework is the same as in the Anderson problem, the transfer matrix calculation is by no means identical; indeed, in the many body case the energies associated to the different graph vertices are a linear combination of the independent random fields, and are thus correlated. Moreover, the number of paths connecting two sites proliferates with the size of the chain $L$, with a scaling that is faster than exponential. These paths present correlations that are much stronger with respect to the Anderson problem, as we shall discuss in more detail in Sec.~\ref{sec:dompath}.

\subsubsection{Distribution of the wave function amplitudes and critical disorder}

\begin{figure}
\begin{center}

 \includegraphics[width = \columnwidth]{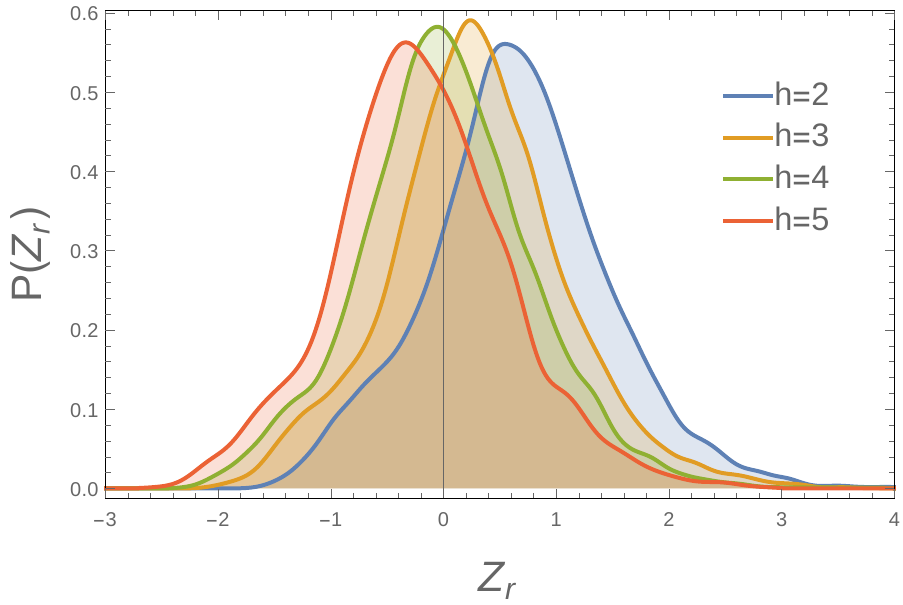}
\caption{(Color online) Probability density of the random variable $Z_r(h)$ defined in Eq.~(\ref{eq:Zrh}), for an $XXZ$ spin  chain of length $L=20$ (corresponding to $r=10$) and different values of disorder $h$. Each curve is obtained with $3 \cdot 10^3$ realizations.}
\label{fig:Pzh}

\end{center}
\end{figure}

\begin{figure}
\begin{center}
 \includegraphics[width = \columnwidth]{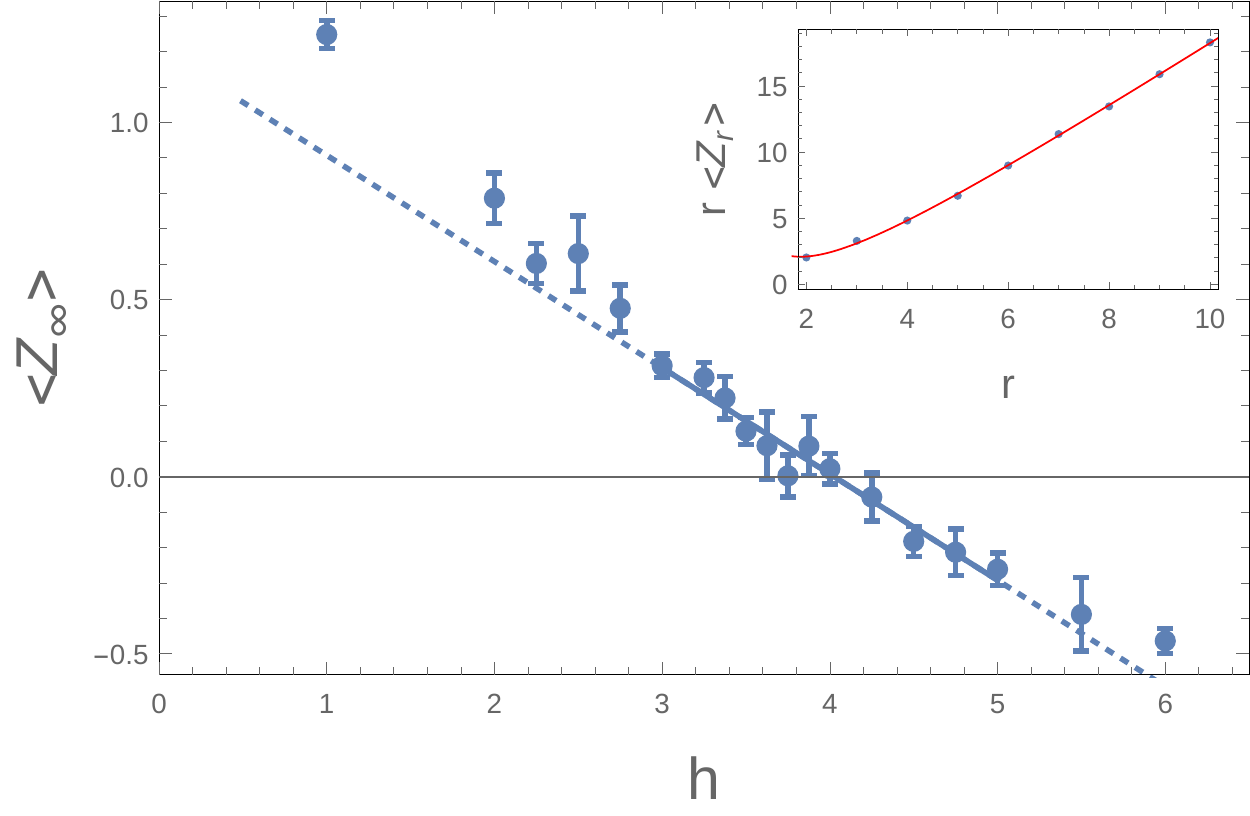}
 \caption{(Color online) Extrapolated value of the mean $\langle Z_\infty\rangle$ of the variable defined in Eq.~(\ref{eq:Zrh}). The
crossing with 0 signals the many body localization/delocalization
transition for $t=1$ (see Eq. \eqref{eq:xxzzmean}). The error bars are obtained from the
fitting procedure (see Inset). The resulting transition value is
$h_c=4.0\pm 0.3$. \textit{Inset. } Finite size scaling of $\langle r\, Z_r \rangle$ with the distance $r$ between the Neel states $n_1$
and $n_2$. The plot corresponds to $h=1$. The fit is linear with an $r^{-1}$ correction, see Eq. \eqref{eq:xxzfinitesize}, with
parameters $c_1=-7.2\pm0.4$, $\langle Z_\infty(2)\rangle=1.23\pm0.02$ and
$c_2=8.8\pm0.7$. The finite-$r$ values for the mean are obtained over
at least $10^4$ realizations for $r<7$ and at least $2\cdot 10^3$
realizations for $r\geq7$.} \label{fig:meanxxz}
\end{center}
\end{figure}

In Fig.~\ref{fig:Pzh} we show the probability density of $Z_r(h)$ for a chain of length $L=20$ and different values of $h$. Since in the many body case it is not possible to simplify the dependence on the disorder strength $h$, the criterion for the transition reads 
\begin{equation}
 \langle Z_\infty(h_c)\rangle=-\log t,
 \label{eq:xxzzmean}
\end{equation}
where $\langle Z_\infty(h)\rangle$ is the  extrapolated value of the average of \eqref{eq:Zrh} for fixed $h$.  Plots of $\langle Z_\infty(h)\rangle$ are given in Fig.~\ref{fig:meanxxz}. Here $\langle Z_\infty(h)\rangle$  is extrapolated from the finite size values using the fitting function
\begin{equation}
 r\,\langle Z_r(h)\rangle=c_1+\langle Z_\infty(h)\rangle\, r+c_2\,r^{-1}.
 \label{eq:xxzfinitesize}
\end{equation}
For $t=1$, the critical point $h_c$ is estimated from the condition $\langle Z_\infty(h_c)\rangle=0$. The resulting value is $h_c=4.0 \pm 0.3$, which is, as expected, larger than the result derived with exact diagonalization. Notice also that the corrections $\propto r^{-1}$ are consistent with the intuition that the $\omega_{FA}=0$ is the correct mean-field scaling for MBL.

 As we discuss in Sec.~\ref{sec:dompath}, in the many body case the sum \eqref{eq:FAxxz} is no longer dominated by a single path. Thus, the algorithm for the best path is not applicable in this context, and the limited system sizes accessible with the transfer matrix do not allow to investigate whether a scaling form exists also for  \eqref{eq:Zrh} in the limit of large $r$. For the available system sizes, the distributions of the rescaled variables $\tilde Z_r(h)=\tonde{Z_r(h)-\langle Z_r(h)\rangle}/{\sigma_{Z_r(h)}}$ do not seem to collapse to a unique curve, and the scaling of the variances $\sigma^2_{Z_r(h)}$ with $r$ appears to be compatible with a power-law, but with exponent depending on the disorder strength $h$. However, a more refined numerical analysis is necessary to draw a conclusion on the asymptotic behavior.

\subsubsection{Divergent length scales and critical exponents}

\begin{figure}
\begin{center}
 \includegraphics[width = \columnwidth]{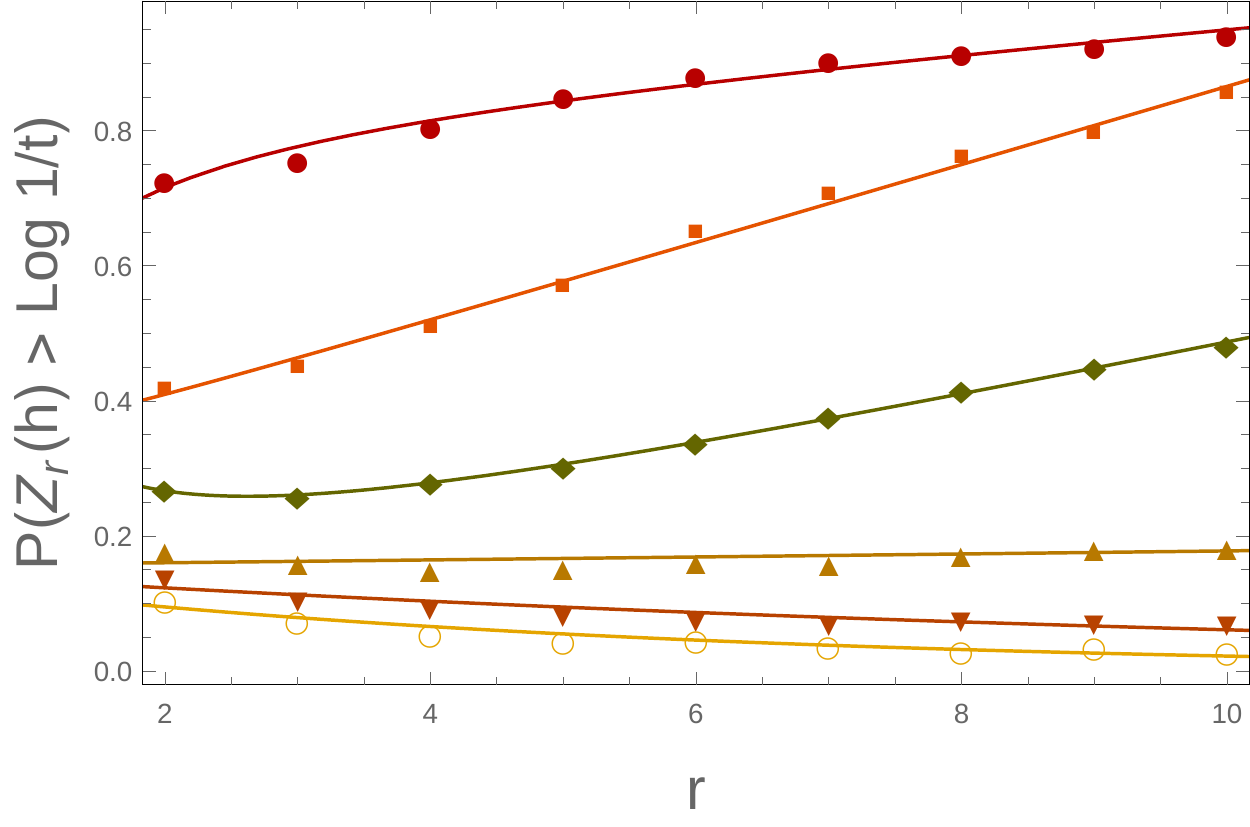}
 \caption{(Color online) Probability of resonances $P(Z_r(h)>-\log t)$ as a function
of the distance $r$ between the two Neel states $n_1$ and $n_2$, for
$t=1$. Asymptotically, the probability reaches zero exponentially in the localized
phase and one in the delocalized phase. We average over $10^4$, $5\cdot 10^3$
and $3\cdot 10^3$ realizations for $r\leq8$, $r=9$ and
$r=10$, respectively; the plotted values of the disorder $h$ are: $h=1$ (points), $h=2$ (squares), $h=3$ (diamonds), $h=4$ (upward triangle), $h=5$ (downward triangle) and $h=6$ (circle). Linear and exponential fits in the delocalized and localized regions respectively are plotted as continuous lines, see Eqs. \eqref{eq:fitdelocMBL} and \eqref{eq:fitlocMBL}.}  \label{fig:xxz}
\end{center}
\end{figure}

Fig.~\ref{fig:xxz} shows the behavior of the probability of resonances $ P \tonde{Z_r(h)> - \log t}$ as function of the distance between the Neel states. As expected, the $r$-dependence  changes with the disorder: the probability decays to zero at large $h$, and increases towards one for the smaller $h$.  We expect the convergence to be exponential in $r$ on both sides of the transition; however, the exponential behavior is not clearly detectable in the delocalized phase, due to the few accessible system sizes. For $h<h_c$ we extract a length scale $l(h)$ by fitting the curves in Fig.~\ref{fig:xxz} with the function:
\begin{equation}\label{eq:fitdelocMBL}
 P \tonde{Z_r(h)> 0}= a_2(h) + \frac{r}{l(h)}+ \frac{b(h)}{r}.
\end{equation}

\begin{figure}
\begin{center}
 \includegraphics[width = \columnwidth]{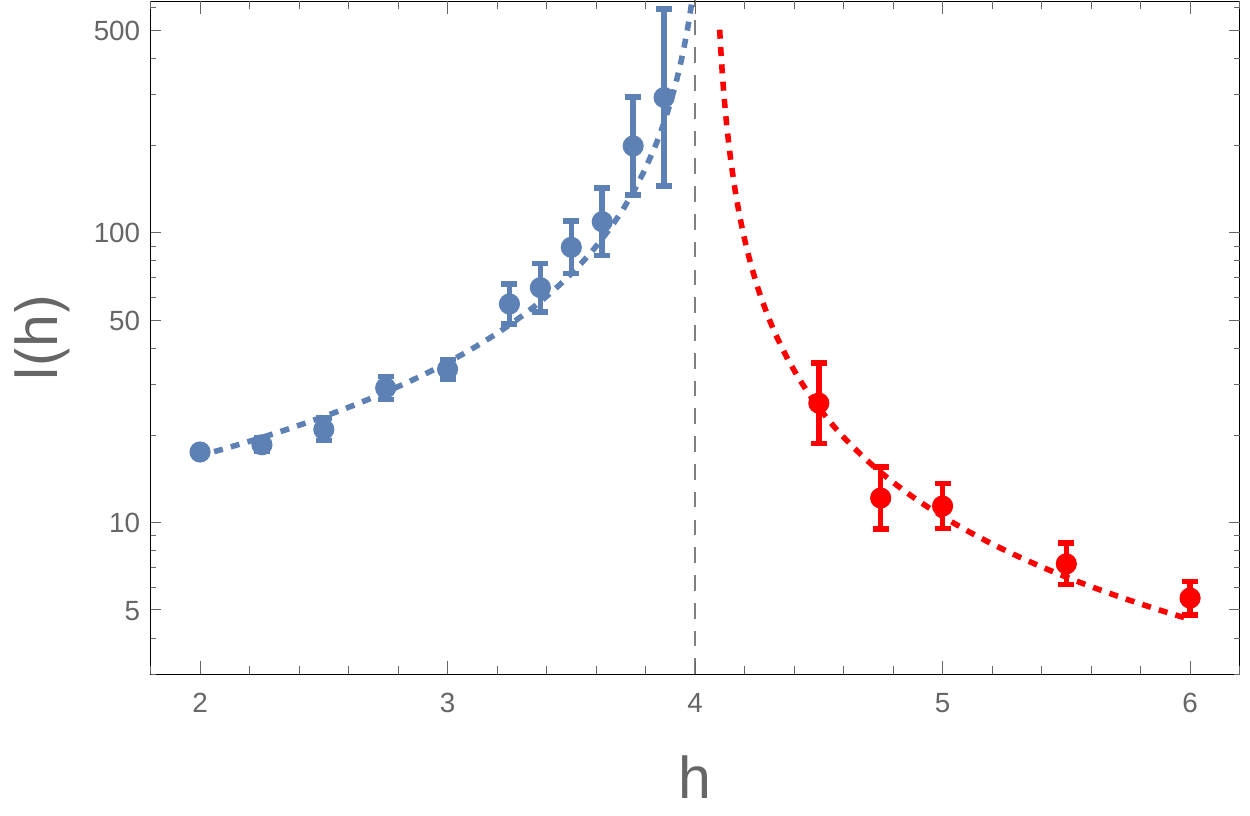}
 \caption{Color online) Divergence of the length scales $l(h)$ extracted from the fits of the probability of resonances as a function of $r$. The vertical dashed line indicates the critical value $h_c$ obtained in Fig.~\ref{fig:meanxxz}. The dotted curve is a power law fit of the form $c |h-h_c|^{-\nu}$, resulting in a critical exponent $\nu_L=1.12 \pm 0.06$ for $h<h_c$ and $\nu_R=1.1 \pm 0.2$ for $h>h_c$.
}\label{fig:xixxz}
\end{center}
\end{figure}

In the localized phase we perform instead the exponential fit:
\begin{equation}\label{eq:fitlocMBL}
P \tonde{Z_r(h)> - \log t}= a_1(h) \text{exp}\tonde{-\frac{r}{l(h)}}.
\end{equation}
 The length scales $l(h)$ extracted with this procedure are shown in Fig.~\ref{fig:xixxz}, together with the power law fit $l(h)= c |h-h_c|^{-\nu}$. The fit is performed separately for $h<h_c$ and $h>h_c$, resulting in an exponent close to $1$ in both cases (see Fig.~\ref{fig:xixxz} for details). Note the asymmetry of the curve with respect to $h_c$, which indicates that at fixed $|h-h_c|$ the typical distance to find a resonance in the delocalized phase is larger than the localization length at the corresponding value of disorder in the localized phase. A possible consequence of this phenomenon, which occurs also in the Anderson model (see Fig.~\ref{fig:nu3}), could be a large ``critical region" in the dynamics in the delocalized phase.

\section{The structure of the dominating paths}
\label{sec:dompath}

As mentioned in section Sec.~\ref{par:hypercube}, the wave function amplitudes in FA can be interpreted as the partition function for a directed polymer in random medium. However, a relevant difference is that while the weight associated to the polymer is bounded from above\cite{deridda1988polymer}, the single factors in \eqref{eq:FAcubeRescaled} are unbounded, with diverging average. As a consequence, the strongly localized phase in the Anderson model (where the FA is best controlled) always corresponds to a ``frozen" phase of the directed polymer, in which most of the weight in the total sum \eqref{eq:forwardWF} is given by one single path. An interesting question\cite{Biroli2012} is whether the freezing phenomenon persists in the delocalized phase in some form, and what this implies that the structure of the eigenstates close to the transition \cite{de2013support,de2014anderson}.

In this section, we compare the statistics of the wave function amplitudes in FA with that of the optimal path, i.e. the path with maximal amplitude, both for the single particle problem and for the XXZ chain. We show that while in the first case the full sum is strongly dominated by the extremal path amplitude, in the many body case most of the paths have comparable amplitude. However, the much stronger correlation between them gives rise to non-negligible interference effects, resulting in strong cancellations.

\subsection{The single particle case}
\begin{figure}
\begin{center}
 \includegraphics[width = \columnwidth]{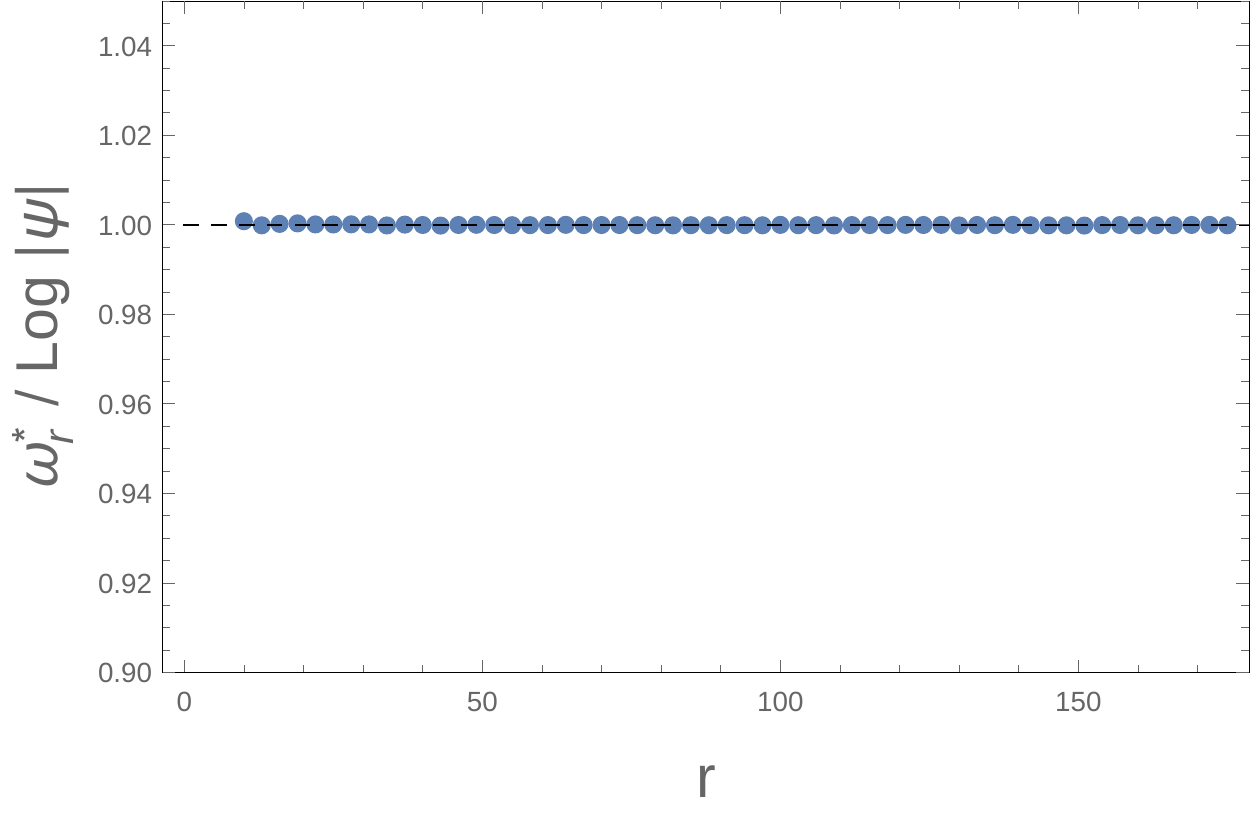}
\caption{Average ratio between the dominating path weight $\omega_r^*$ computed with the Dijkstra algorithm, and the sum in Eq. \eqref{eq:FAcube} computed using a transfer matrix technique. The ratio is taken between values corresponding to the same disorder realization. The plot corresponds to $d=3$, and each point is averaged over $3\cdot 10^4$ disorder realizations. Similar results are obtained for higher dimensionality, for those $r$ accessible with the transfer matrix technique. The standard deviation error bars are within the point size. }\label{fig:Ratio}
\end{center}
\end{figure}

For the finite dimensional case, we compute the amplitude $\omega^*_r$ of the optimal path $p^*$ dominating the sum \eqref{eq:forwardWF} by means of the Dijkstra algorithm \cite{dijkstra1959}, a graph-search algorithm that determines the path minimizing a given cost function. We consider a directed cube (such as the one in Figure \ref{fig:hypercube}) and assign a positive cost $\chi$ to each directed edge $\langle i, j \rangle$:
\begin{equation}
\label{eq:weightsSP}
 \chi \tonde{i,j} \equiv \log |\epsilon_j| - \min_k\left\{ \log |\epsilon_k | \right\}.
\end{equation} 
The total cost of a path $p$ is the sum of the costs of the edges belonging to it, and the path $p^*$ with maximal amplitude is the one minimizing the total cost function. In order to compare with the transfer matrix results, we compute the ratio between $\omega_r^*$  and the full sum \eqref{eq:FAcube} computed via the transfer matrix technique, for the same given disorder realization. The distribution of the ratios turns out to be very narrowly peaked around one.  Figure \ref{fig:Ratio} displays its average as a function of the length of the paths $r$ for $d=3$, which is extremely close to one, uniformly in the path length.

As a further check of the agreement between the values computed with the two methods, we plot in Fig.~\ref{fig:PLfit} the $r$-dependence of the probability \eqref{eq:pres} with $\delta=0$, determined with the substitution $|\psi_r| \to \omega^*_r$. The data are plotted as points, which are almost indistinguishable from the transfer matrix results (squares). This indicates that the statistics of distant resonances is fully captured by the optimal path. Thus, in the single particle case the correlation between different paths does not play a relevant role, since the sum is dominated by the extremum, as it would happen for independent random variables with fat-tailed distribution. Based on this observation, the numerical analysis outlined in the previous sections can be carried out for much bigger system sizes with respect to the ones accessible with the transfer matrix technique, since the Dijkstra algorithm has lower complexity than the transfer matrix (indeed, the time complexity is $\sim O(e+v\log v)$ and the space complexity is $\sim O(v^2)$, where $v$ is the number of vertices and $e$ is the number of graph edges).

We perform the same analysis also for the modified forward approximation discussed in Ref.~\onlinecite{Somoza2009}, by taking the energy denominators uniformly distributed in $\quadre{-1, -W^{-1}} \cup \quadre{W^{-1}, 1}$ in $d=3$ for two values of the cutoff, $W=25$ and $W=35$. We find that in this case the ratio between the maximal path and the transfer matrix result departs from one for increasing $r$. This suggests that more than one path dominates the transfer matrix result. It is natural to expect that in this case the number of dominating paths depends on the geometry of the system, thus introducing a stronger dependence on the dimensionality, see also the comments in Sec.~\ref{par:hypercube}.

\begin{figure}
\begin{center}
 \includegraphics[width = \columnwidth]{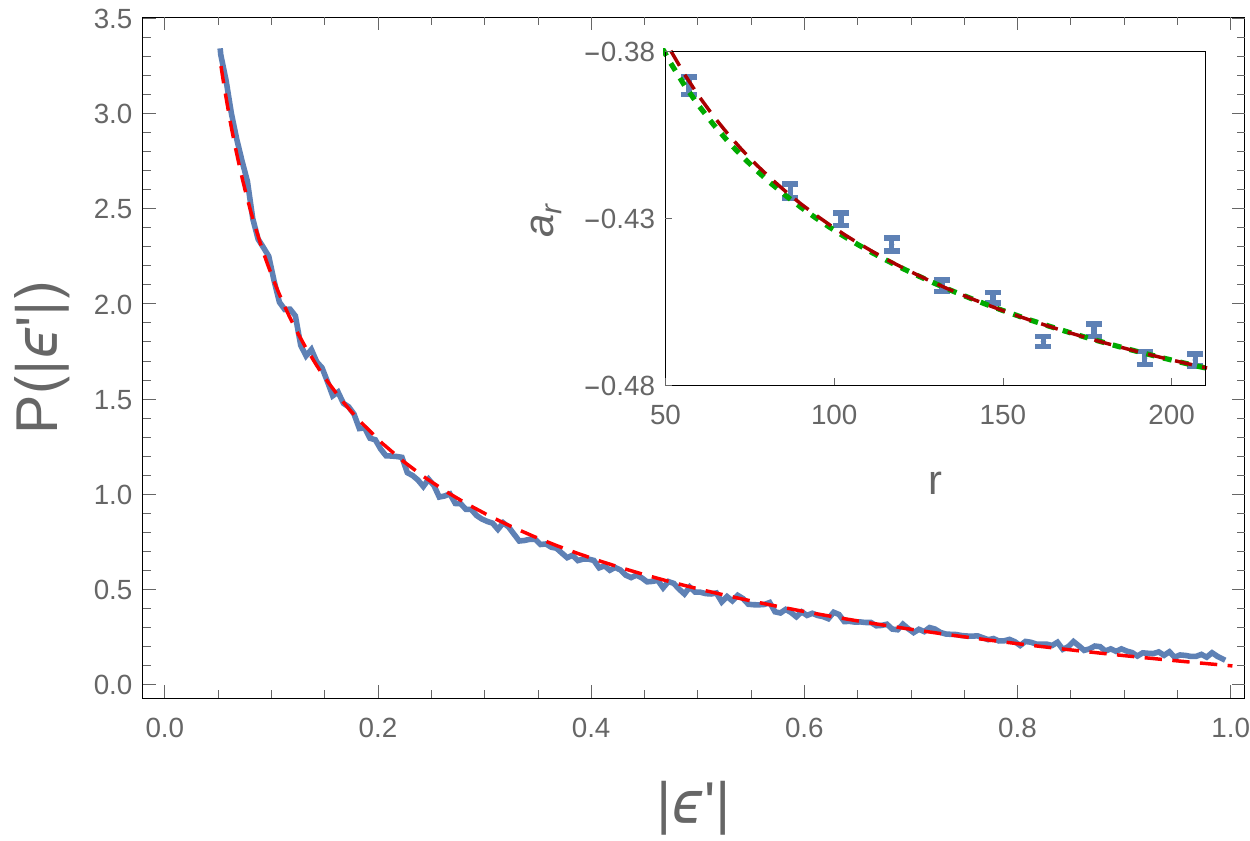}
 \caption{(Color online). Probability distribution $\rho(\epsilon)$ of the energy denominators along the optimal path, see Eq.~\eqref{eq:Yu1}. The plot corresponds to $d=3$ and $r=210$. The dashed red line is the fitting function of Eq.~\eqref{eq:fitRho}, with fitting parameters $c_r=-0.95\pm0.04$, $b_r=1.04\pm0.03$ and $a_r=0.472\pm 0.005$. Very similar results are obtained for higher dimensionality. \textit{Inset. } Plot of the exponents $a_r$ in the fitting function of Eq.~\eqref{eq:fitRho}, as a function of $r$. Due to the absence of a theoretical reasoning for the finite size scaling, we fit the curve considering logarithmic and $1/\sqrt{r}$ corrections. The green small-dashed curve is a fitting function of the form $a+ c/ \log (r)$, with fit parameters $a=-0.73\pm0.05$ and  $c=-1.4\pm0.3$; 
 the red large-dashed curve is a fitting function of the form $a+c/\sqrt{r}$, with fit parameters $a=-0.57\pm0.02$ and $c=-1.4\pm0.3$.
 The asymptotic value $a$ obtained with the logarithmic fitting function is compatible  with the solution of the equation \eqref{eq:Yu2} for $d=3$. 
  }\label{fig:EnergyDist}
\end{center}
\end{figure}

For the case of unbounded denominators, we compute the inverse participation ratio (IPR) of the edge weights contributing to  $\omega^*_r$, for $\epsilon_i \in \quadre{-1,1}$ (i.e. $W=2$). We define
\begin{equation}
\label{eq:IPR}
 \text{IPR}=\frac{\left( \sum_i \log|\epsilon_i|\right)^2}{\sum_i (\log |\epsilon_i|)^2},
\end{equation}
where $i$ labels the sites belonging to the optimal path $p^*$. We find that the disorder-averaged IPR scales linearly with the length of the path $r$, indicating that an extensive (in $r$) number of edges contributes to the total path weight, and cooperate to produce the atypically big path weights dominating \eqref{eq:FAcube}. 
Fig.~\ref{fig:EnergyDist} shows the distribution of the absolute value of the energies along the optimal path for $W=2$, $d=3$, $r=210$ and $\epsilon_a=0$. The fitting function has the form 
\begin{equation}\label{eq:fitRho}
\rho_r(\epsilon)=c_r +b_r |\epsilon|^{a_r}.
\end{equation}
The power-law behavior is consistent with the considerations in Ref.~\onlinecite{Yu2013}. Adapting their reasoning to the finite dimensional case, one can argue that asymptotically in $r$ (and under the hypothesis of independent paths) the biased energy distribution along the optimal path has the form
\begin{equation}
\label{eq:Yu1}
 \rho(\epsilon)= \frac{1-2x}{|\epsilon|^{2x}},
\end{equation}
with $x$ solving the $d$-dependent equation
\begin{equation}
\label{eq:Yu2}
 \log \tonde{\frac{d}{1-2x}}- \frac{2x}{1-2x}=0.
\end{equation}
Fitting the $r$-dependence of the coefficients $c_r, b_r, a_r$ one finds that the asymptotic limits are in agreement with \eqref{eq:Yu1}; for details see the inset of Fig.~\ref{fig:EnergyDist}.

\begin{figure}
\begin{center}
 \includegraphics[width = \columnwidth]{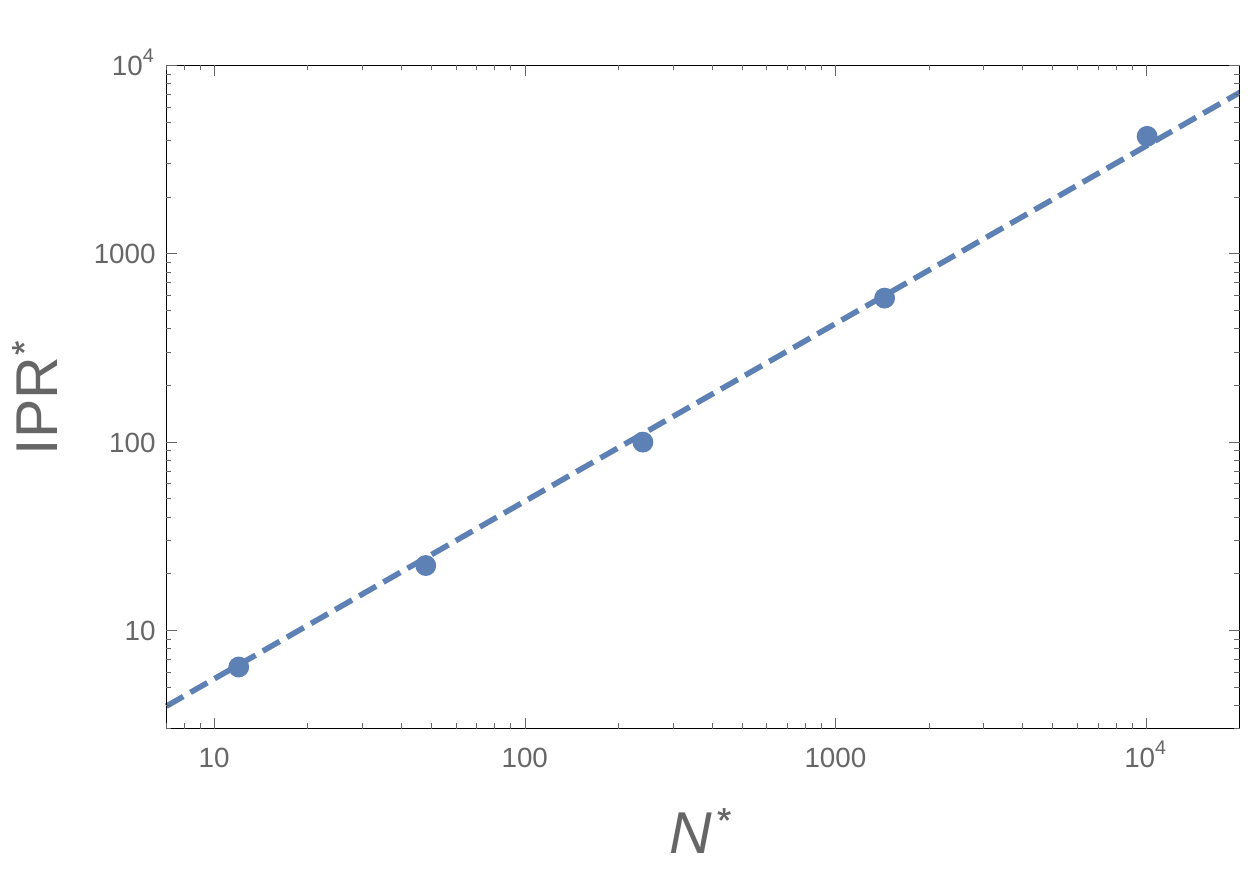}
\caption{Average IPR* of the paths, Eq.~(\ref{eq:IPR2}),  as a function of the number of paths $N^*= 2 (L/2)!$ for random spin chains of different lengths $L$. The IPR* is linear in the total number of paths. 
}\label{fig:iprpaths}
\end{center}
\end{figure}

\subsection{The many-body case}
When performing the same type of analysis for the Heisenberg chain, we find that the statistics of the sum \eqref{eq:FAxxz} is not well reproduced by the optimal path alone: the distribution of the ratios between the full sum and the optimal path is very wide and peaked at values that are far from one. Thus, despite also in this case the amplitude of the single paths are fat-tailed distributed, there is not a single one dominating. Instead, we find that the average IPR* of the paths amplitudes (which we denote with $\omega_p$):
\begin{equation}
\label{eq:IPR2}
 \text{IPR*}= \frac{\tonde{\sum_p \omega_p}^2}{\sum_p \omega_p^2}
\end{equation}
scales linearly with the total number of paths $N^*=2(L/2)!$, indicating that there are factorially many (in the length of the chain $L$) paths having amplitudes that are comparable in absolute value. This is a signature of the strong correlations between the paths, which is not surprising in view of the many-body nature of the model. Following Ref.~\onlinecite{ros2015integrals}, one can argue that the strongest correlations are among those paths associated to processes in which the same spin flips occur, but in different order: the different orderings of the flips produce different energy denominators in \eqref{eq:FAxxz}, and thus different path amplitudes; however, the resulting terms are correlated, and one can expect that for those realizations of the random fields producing one particularly large path weight, the other ones (related to it by permutation of the order of the spin flips) will also have a large amplitude in absolute value. However, in the sum \eqref{eq:FAxxz} the paths contribute with well defined relative signs, leading to cancellations between these factorially many terms (see Ref.~\onlinecite{ros2015integrals} for an explicit calculation for a model of interacting fermions), which are fully taken into account only with the transfer matrix method.

\section{Conclusion}
In this work we have discussed the advantages and the limitations of the forward approximation applied to both single- and many-body quantum disordered systems. In particular, the FA has been used to obtain an expression for the wave functions, that can be computed by means of a transfer matrix technique. The statistical analysis of the wave functions allows to determine the critical values of the disorder (exact in large $d$), the critical exponents of the localization length (which turn out to be mean-field) and the universal distribution of the eigenfunctions' coefficients. 

For the single particle case, the amplitudes of the wave functions in FA turned out to be very well approximated by only one path, the dominating path: this has been exploited to investigate larger system sizes with respect to the ones accessible with the transfer matrix (the algorithm that computes the best path runs in time linear in the number of edges of the underlying graph, and is not as memory-demanding as either the transfer matrix of shift-invert exact diagonalization). The extremely good agreement (within statistical error) of the predicted critical values for the disorder for $d\geq 5$ suggests that the approximation should be predictive also for the properties of the wave functions at the critical point. We have not investigated in details these implications, but we foresee the wave functions to have a sparse structure which is similar to that discussed in high-coordination Bethe lattices\cite{de2014anderson}. The strong similarities with the problem of directed polymers in random medium have also been addressed; however, from the statistical analysis it emerges that the scaling exponents describing the fluctuations of the wave functions are non-mean field, but also not equal to those of the directed polymer. Moreover, the limiting distribution of the appropriately rescaled wave functions seems to depend weaker on the dimensionality with respect to the directed polymer case.

In many-body problem there is no concentration of amplitude on a small number of paths, but there are strong cancellations between them: as a result, the full sum over factorially-many paths is only exponentially large (or small) in the system size. The correlation between the paths has been discussed in detail in Ref.~\onlinecite{ros2015integrals}, and this work can be interpreted as a numerical test of the claims in that work. 
For the XXZ chain with random fields, the critical value predicted within the approximation ($h_c=4.0\pm 0.3$) is very close to the most updated result obtained with exact diagonalization; thus, this tool furnishes an alternative route to exact diagonalization, that can be applied to significantly larger system sizes. We leave to future work the question of how to incorporate higher-order corrections in the FA within the transfer matrix scheme, and how they affect the critical exponents, and the accuracy of the results.

As a conclusive remark, we would like to briefly comment on the nature of the FA as a mean-field approximation for Anderson localization. For certain, the value of the critical disorder $W_c$ for the onset of localization grows indefinitely with $d$, and in high $d$ the hopping $t$ becomes an almost negligible perturbation at the transition. The fact that the error in $W_c$ essentially disappears around $d\simeq 6$ is a strong indication of this. This feature is quite peculiar, since in ordinary, second order phase transitions the critical exponents above the upper critical dimension are correctly reproduced by the mean field approximation, but the location of the transition (e.g.\ the critical temperature) is not. In this sense the locator expansion (that to lowest order reduces to the FA considered in this work) becomes a better suited candidate for a mean field than the $2+\epsilon$ expansion of the nonlinear supersymmetric sigma model (NLS$\sigma$M)\cite{efetov1999supersymmetry}. It is plausible that there is a field-theoretical description of the FA which can be put in direct relation with the NLS$\sigma$M; this is an obvious direction in which to continue this work. In addition, the relation with the Bethe lattice results can be further investigated, given that the critical $W_c$ predicted by the FA does not correspond to that of a Bethe lattice of any (integer) coordination number.

\section{Acknowledgements}
AS would like to thank B.L.Altshuler, R.Moessner, V.Oganesyan and S.L.Sondhi for discussions.

\appendix
\section{Probability density of $Z_r$}\label{appendix:ProbDen}

An estimate for the probability density of $Z_r$ can be obtained making use of the fact that the sum over \eqref{eq:FAcubeRescaled} is dominated by the single path with maximal weight:
\begin{equation}
 \label{eq:zAsMax}
 Z_r \approx \max_{p\in\text{spaths}} \grafe{-\frac{1}{r} \sum_{i \in p} \log |\epsilon_i'|}.
\end{equation}

If the correlations between the different path weights are neglected, the calculation is similar to the one performed for the Bethe lattice case. In particular, one finds for the cumulative function of $Z_r$ the following expression:
\begin{equation}
\label{eq:cumulative}
 P \tonde{Z_r < a}= \text{exp}\mkern-4mu\quadre{ \mkern-3mu N_r \log \mkern-3mu \tonde{\mkern-3mu 1-\frac{1}{(r-1)!} \int_{r(a-\log2)}^{\infty} \mkern-15mu t^{r-1} e^{-t} dt \mkern-3mu}\mkern-3mu},
\end{equation}
where $N_r \sim d^r$ is the total number of paths on which the maximum is taken. This implies the following form for  the probability density of $Z'_r=Z_r-\log 2$:
\begin{equation}
 \label{eq:DistZ}
 p_r(z')= \frac{N_r r^r}{(r-1)!} \frac{e^{- r \tonde{z' -\log z' } }}{z'  \quadre{1-I_r(z')}} \text{exp} \quadre{N_r \log \tonde{1-I_r(z')}} ,
\end{equation}
where we introduced the monotone decreasing function
\begin{equation}
\label{eq:intProb}
 I_r(z')= \frac{1}{(r-1)!} \int_{r z'}^{\infty} t^{r-1} e^{-t} dt.
\end{equation}

The typical value of $Z'_r$, denoted $z^*_r$, is defined by the equation
\begin{equation}
\label{eq:critTyp}
 N_r \, I_r(z^*_r)=N_r \frac{r^{r-1}}{(r-1)!}\int_{z^*_r}^\infty t^{r-1}e^{-r t}dt= 1.
\end{equation}

The solutions of \eqref{eq:critTyp} approach a finite limit $z^*$ for $r \to \infty$, which is related to the critical value of disorder by $z^*= \log (W_c/(2 t))$, as previously discussed. Using that $N_r \sim d^r$ and computing the integral in \eqref{eq:critTyp} with a saddle point calculation (assuming $z^* >1$), one recovers the condition \eqref{eq:condAbu} for $W_c$, with the substitution $K \to d$.

For increasing $r$ the probability density of $Z'_r$ peaks at the typical value, with tails that approach zero in the limit $r \to \infty$. In particular, for $z' >z^*_r$ the decay of the tail is exponential in $r$. Indeed, in this regime the product $N_r \, I_r(z)$ is itself exponentially decreasing with $r$; thus, the rightmost exponential in \eqref{eq:DistZ} rapidly converges to one, and the distribution $p_r(z')$ approaches zero with a tail of the form
\begin{equation}
\label{eq:tail1}
  p_r(z') \sim e^{- r \tonde{z' -\log z'- \log (d e)}+ o(r)}.
\end{equation}
When $z'$ becomes smaller than the typical value $z^*_r$, the product $N_r \, I_r(z')$ increases exponentially. Since for large $r$ the integral  $I_r(z')$ is still exponentially small for all $z' > 1 +O(1/r)$, one can still set $\log \quadre{1- I_r(z')} \sim - I_r(z') \sim \text{exp}\tonde{-r z'+r \log(e z')+o(r)}$. Thus, in this regime the probability density of $Z'_r$ decays to zero much faster, double-exponentially with $r$
\begin{equation}
\label{eq:tail2}
 p_r(z') \sim \text{exp}\tonde{- d^r \; e^{-r z'+r \log(e z')} + O(r)}.
\end{equation}
 Note that (for $r$ large enough) the interval in which $1 < z'< z^*_r$ does not shrink to zero for $d \geq 3$, given that the value $z^*$ obtained from the condition \eqref{eq:critTyp} is always bigger than one. When $z'$ approaches one, the probability in \eqref{eq:intProb} is no longer a large deviation probability, i.e. it is no longer exponentially small in $r$: the term $\log \quadre{1-I_r(z')}$ approaches a constant function of $z'$, and the main scaling is given by the factor $d^r$.

Finally, exactly at $z'=z^*_r$, using \eqref{eq:critTyp} and performing the integral with an integration by parts, one finds that the probability density can be written as
\begin{equation}
\begin{split}
 \label{eq:delta}
  p_r(z^*_r)= &\frac{r}{1-d^{-r}} \quadre{1 + \sum_{n=1}^{r-1} \frac{(r-1)!}{(r-1-n)! r^n} (z^*_r)^{-n}}^{-1}  \times \\
  &\text{exp} \tonde{-1 - \frac{1}{2 d^r}- \frac{1}{3 d^{2r}}+ \cdots},
  \end{split}
\end{equation}
which diverges like $r$ when $r \to \infty$.

Given the tails of the distribution of $Z_r$ computed in this approximation, it is immediate to derive the asymptotic decay of the probability of resonances in the localized phase. Indeed, for $W>W_c$, the probability \eqref{eq:presResc} is is a large deviation for $Z_r$. Making use of \eqref{eq:tail1} we find
\begin{equation}
\begin{split}
 \label{eq:maxPath}
P \tonde{Z_r > \log \frac{W}{t}}&= \int_{\log\tonde{\frac{W}{2t}}}^{\infty} e^{- r \tonde{z' -\log z'- \log (d e)}+ o(r)} dz'\\
&= \text{exp}\tonde{- \frac{r}{l(W)}+ o(r)}
\end{split}
\end{equation}
with 
\begin{equation}
 \frac{1}{l(W)}= \log \tonde{\frac{W}{2 t d e} \frac{1}{\log \tonde{W/2t}}}. 
\end{equation}

Thus, within this approximation for $W$ approaching $W_c$ from above one finds 
\begin{equation}
 l(W) \sim \frac{W_c}{W-W_c},
\end{equation}
thus the length scale diverges at the transition with a critical exponent equal to $1$.

Similarly, for $2 t e < W< W_c$, making use of \eqref{eq:cumulative} and of \eqref{eq:tail2} we find:
\begin{equation}
\begin{split}
 P \tonde{Z_r < \log \tonde{\frac{W}{t}}}&\approx \text{exp}\tonde{- \quadre{\frac{2 t d e}{W} \log \tonde{\frac{W}{2t}}}^r+ O(r) }\\
 &=\text{exp}\tonde{- e^{-r/l(W)}+ O(r) }.
 \end{split}
\end{equation}

\bibliography{biblioMBL}
 \end{document}